\documentclass[
 aps,
 prl,
 reprint,
 amsmath,
 amssymb,
 superscriptaddress,
 nofootinbib
]{revtex4-2}

\usepackage{bm}
\usepackage{dcolumn}
\usepackage{graphicx}
\usepackage{hyperref}
\usepackage{units}
\usepackage{xcolor}

\begin{document}
\title{Role of chiral two-body currents in $^6$Li magnetic properties in light of a new precision measurement with the relative self-absorption technique}
\date{\today}
\newcommand{\emmi}{\affiliation{ExtreMe Matter Institute EMMI, GSI Helmholtzzentrum f\"ur Schwerionenforschung GmbH, 64289 Darmstadt, Germany}}
\newcommand{\gsi}{\affiliation{GSI Helmholtzzentrum f\"ur Schwerionenforschung GmbH, 64289 Darmstadt, Germany}}
\newcommand{\ikpjguprisma}{\affiliation{Institut f\"ur Kernphysik and PRISMA Cluster of Excellence, Johannes Gutenberg-Universit\"at Mainz, 55128 Mainz, Germany}}
\newcommand{\mainzgsi}{\affiliation{Helmholtz Institute Mainz, GSI Helmholtzzentrum f\"ur Schwerionenforschung GmbH, 64289 Darmstadt, Germany}}
\newcommand{\ikptud}{\affiliation{Institut f\"ur Kernphysik, Technische Universit\"at Darmstadt, 64289 Darmstadt, Germany}
}
\newcommand{\mpi}{\affiliation{Max-Planck-Institut f\"ur Kernphysik, 69117 Heidelberg, Germany}}
\newcommand{\mstud}{\affiliation{Department of Materials Science, Technische Universit\"at Darmstadt, 64287 Darmstadt, Germany}
}
\newcommand{\nsu}{\affiliation{Department of Physics, North Carolina State University, Raleigh, NC 27695, USA}}
\newcommand{\supa}{\affiliation{SUPA, Scottish Universities Physics Alliance, Glasgow, G12 8QQ, UK}}
\newcommand{\triumf}{\affiliation{TRIUMF, 4004 Wesbrook Mall, Vancouver BC, V6T 2A3, Canada}}
\newcommand{\tunl}{\affiliation{Triangle Universities Nuclear Laboratory, Duke University, Durham, NC 27708, USA}}
\newcommand{\ubc}{\affiliation{Department of Physics and Astronomy, University of British Columbia, Vancouver BC, V6T 1Z4, Canada}}
\newcommand{\unc}{\affiliation{Department of Physics and Astronomy, University of North Carolina at Chapel Hill, Chapel Hill, NC 27599, USA}}
\newcommand{\uom}{\affiliation{Department of Physics and Astronomy, University of Manitoba, Winnipeg MB, R3T 2N2, Canada}}
\newcommand{\uws}{\affiliation{School of Engineering, University of the West of Scotland, Paisley, PA1 2BE, UK}}

\author{U.~Friman-Gayer} \email{ufrimangayer@ikp.tu-darmstadt.de} \ikptud \unc \tunl
\author{C.~Romig} \altaffiliation[Present address: ]{Projekttr\"ager DESY, Deutsches Elektronen-Synchrotron, 22607 Hamburg, Germany} \ikptud

\author{T.~H\"uther} \ikptud
\author{K.~Albe} \mstud
\author{S.~Bacca} \ikpjguprisma \mainzgsi
\author{T.~Beck} \ikptud
\author{M.~Berger} \ikptud
\author{J.~Birkhan} \ikptud
\author{K.~Hebeler} \ikptud \emmi
\author{O.~J.~Hernandez} \ubc \ikpjguprisma
\author{J.~Isaak} \ikptud
\author{S.~K\"onig} \ikptud \emmi \nsu
\author{N.~Pietralla} \ikptud
\author{P.~C.~Ries} \ikptud
\author{J.~Rohrer} \mstud
\author{R.~Roth} \ikptud
\author{D.~Savran} \gsi
\author{M.~Scheck} \ikptud \uws \supa
\author{A.~Schwenk} \ikptud \emmi \mpi
\author{R.~Seutin} \mpi \ikptud \emmi
\author{V.~Werner} \ikptud

\begin{abstract}
	A direct measurement of the decay width of the excited $0^+_1$ state of $^6$Li using the relative self-absorption technique is reported.
Our value of $\Gamma_{\gamma, 0^+_1 \to 1^+_1} = \unit[8.17(14)_\mathrm{stat.}(11)_\mathrm{syst.}]{eV}$ provides sufficiently low experimental uncertainties to test modern theories of nuclear forces.
The corresponding transition rate is compared to the results of \textit{ab~initio} calculations based on chiral effective field theory that take into account contributions to the magnetic dipole operator beyond leading order.
This enables a precision test of the impact of two-body currents that enter at next-to-leading order.
\end{abstract}

\maketitle

Nuclear structure physics has entered an era of precision studies, both in experiment and theory.
For light nuclei, \textit{ab initio} theory based on interactions from chiral effective field theory \cite{Epelbaum09} is reaching an accuracy at which corrections to electromagnetic (EM) operators which emerge naturally in the chiral expansion become relevant.
A recent review \cite{Bacca2014} indicates that precision measurements of EM transition rates with uncertainties of a few percent or better are required to explore and validate the effects of these subleading corrections.
For few-nucleon systems, direct measurements of strong transition rates with such precision are often challenging experimentally owing to the very short lifetimes involved.

The present study is focused on the nucleus $^6$Li in its excited $0^+_1$ state at $E_{0^+_1} = \unit[3562.88(10)]{keV}$ \cite{Tilley02}, which constitutes the lightest non-strange hadronic system \cite{Zyla2020} with a dominant internal EM decay branch to its $1^+_1$ ground state.
The potentially competing parity-forbidden decay via $\alpha$ emission has not been observed, and it is at least ten million times weaker than the $\gamma$ decay \cite{Robertson84}.
Because of its occurrence as stable matter (compared to the lighter hypernuclei \cite{Hashimoto06}) and the low nucleon number of $^6$Li, the decay of its $0^+_1$ state is the EM transition of the simplest hadronic system which is simultaneously accessible by precision studies in theory and experiment.
It is, therefore, ideally suited for testing our understanding of nuclear forces and EM currents in a many-nucleon system.
Moreover, this is the EM-analog transition to the $^6$He beta decay, whose rate was recently measured with high precision~\cite{Knecht2012}, so that this $A=6$ system in future offers a comprehensive test of electroweak interactions in light nuclei.

On the theory side, significant progress has been made in chiral effective field theory ($\chi$EFT) \cite{Epelbaum09, EntemMachleidtNosyk2017}, and in the \textit{ab initio} solution of the quantum many-body problem for light nuclei \cite{Barrett13, Carlson15}.
Recently, the focus has been on the consistent inclusion of electroweak transition operators \cite{Bacca2014}, with a focus on the impact of two-body currents (2BC).
For EM transitions in light nuclei, calculations with traditional 2BC and potentials were performed in Ref.~\cite{Marcucci2008}, while calculations with 2BC from $\chi$EFT used in conjunction with wave functions derived from traditional potentials were performed in Ref.~\cite{Pastore13}, reaching a precision at the few-percent level.
In Ref.~\cite{Calci2016}, this transition and the magnetic moment of the ground state of $^6$Li were calculated with interactions from $\chi$EFT without 2BC.
Similarly, these observables have been studied in Refs.~\cite{Parzuchowski2017, Shin2017, Binder2018}.
In this work, we will present the first calculations obtained with 2BC and consistent interactions derived from $\chi$EFT.
In the case of weak $\beta$ decays, this has been shown to lead to a systematic improvement between experiment and theory \cite{Gysbers2019}.

From the experimental side, the determination of the isovector magnetic dipole transition strength
$B(\mathrm{M1}; 0^+_{1, T=1} \to 1^+_{1, T=0}) \propto  E_\gamma^{-3} \Gamma_{\gamma, 0^+_1 \to 1^+_1}$
between the first excited $0^+_1$ state of $^6$Li with a total isospin quantum number of $T=1$ and the $T=0$ ground state, which is proportional to the product of the level width for $\gamma$ decay $\Gamma_{\gamma, 0^+_1 \to 1^+_1}$ and a $\gamma$-ray energy ($E_\gamma$) dependent factor, has been the subject of considerable effort in the past.
The extremely short half-life of the excited state of about \unit[80]{attoseconds ($10^{-18}$s)} \cite{Tilley02} makes a direct measurement of its decay rate impossible \cite{Nolan79}.
\begin{figure*}[!htb]
\includegraphics[width=\textwidth, trim=110 0 110 0, clip]{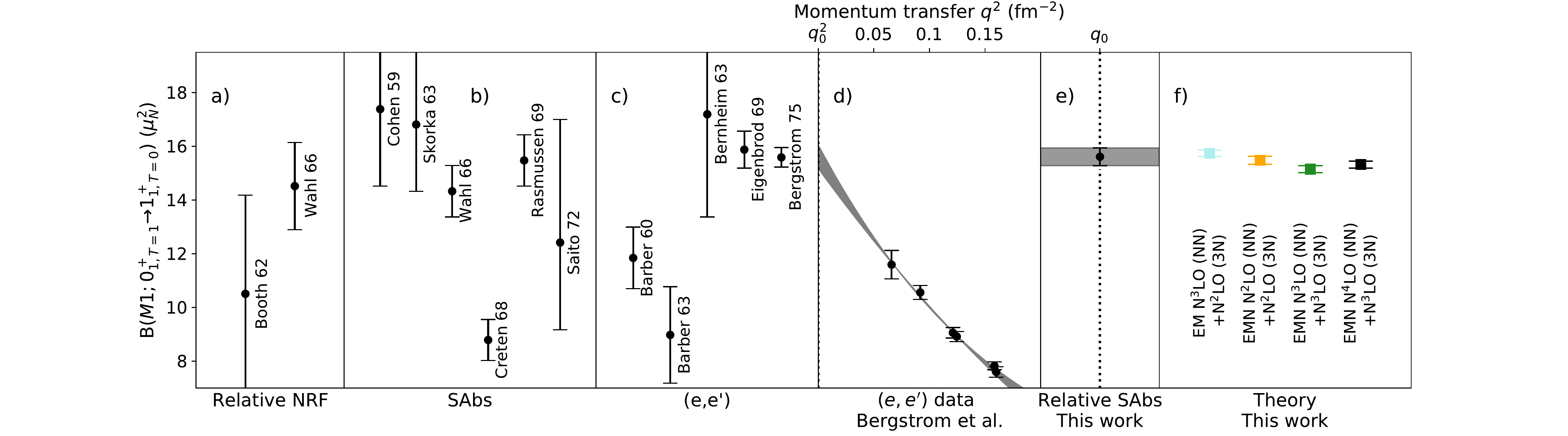}
	\caption{\label{BM1_overview} (a-c): Previous measurements of the $\mathrm{B}(M1; 1^+_1 \to 0^+_1)$ strength for $^6$Li with the methods of relative NRF \cite{Booth62, Skorka66, *Wahl66} (a), SAbs \cite{Cohen59, Skorka63, Skorka66, *Wahl66, Creten68, Rasmussen69, Saito73} (b), and $(e, e')$ \cite{Barber60, Barber63, Bernheim63, Eigenbrod69, Bergstrom75} (c).
	For each experimental method, the data are sorted by the time of publication, with the most recent data point on the right.
	A label next to the data points indicates the year of publication and the last name of the first author.
	Low-$q$ data of the most precise $(e,e')$ result by Bergstrom et al. \cite{Bergstrom75} and a possible quartic polynomial extrapolation (see also the supplemental material \cite{SupplementalMaterialeep}) of $B(M1,q)$ to the photon point ($q_0$) are shown as an uncertainty band in (d).
	The present result, which can be interpreted as a measurement at $q_0$, is given in panel (e).
	Panel (f) shows the result of four theoretical calculations from the present work (see also Fig. \ref{theory_6li}) with estimated uncertainties of the many-body method.
	They employed different Hamiltonians that are indicated by different colors and the labels below the data points and include the leading two-body currents.}
\end{figure*}
Panels (a)-(c) of Fig.~\ref{BM1_overview} show the history of published values for this quantity as compiled in the Evaluated Nuclear Structure Data Files (ENSDF) \cite{Tilley02}.
They have been obtained using three different techniques, namely: nuclear resonance fluorescence relative to another transition (relative NRF) \cite{Booth62, Skorka66, *Wahl66}, self absorption (SAbs) \cite{Cohen59, Skorka63, Skorka66, *Wahl66, Creten68, Rasmussen69, Saito73}, and inelastic electron scattering $(e,e')$ \cite{Barber60, Barber63, Bernheim63, Eigenbrod69, Bergstrom75}.
In the ENSDF, a weighted average value of $B(\mathrm{M1})_\mathrm{ENSDF} = \unit[15.65(32)]{\mu_N^2}$ \{from $\Gamma_{\gamma, 0^+_1 \to 1^+_1} = \unit[8.19(17)]{eV}$ \cite{AjzenbergSelove88, Tilley02}\} is derived from a selection of three of the most recent publications in Ref.~\cite{AjzenbergSelove79}, meant to exclude earlier measurements with obvious systematic deviations (for comparison: a weighted average of all measurements yields a value of $B(M1) = \unit[14.53^{+0.20}_{-0.30}]{\mu_N^2}$).
Regardless of the averaging procedure, the final result is strongly dominated by two $(e,e')$ results of Eigenbrod \cite{Eigenbrod69} and, in particular, of Bergstrom \textit{et al.} \cite{Bergstrom75}, which claim the highest precision.
In such an $(e, e^\prime)$ experiment, the $B(M1)$ value is obtained in a model-dependent way from the measured form factor $\left| F(q) \right|^2$, where $q$ denotes the momentum transfer. 
Both works employed the plane-wave Born approximation to obtain the $q$-dependent $B(M1, q)$ from $\left| F(q) \right|^2$ \cite{Theissen1972}, which is equal to the $B(M1)$ strength in the limit of the minimum necessary momentum transfer $q_0 = \nicefrac{E_{0^+_1}}{\hbar c} \approx \unit[0.018]{fm}$, the so-called 'photon point'.
Panel (d) of Fig. \ref{BM1_overview} shows $B(M1, q)$ values, obtained from the form factor of Bergstrom \textit{et al.} \cite{Bergstrom75}, along with an uncertainty band from one possible extrapolation in our attempt to reproduce their results \cite{SupplementalMaterialeep}.
Similar to Refs. \cite{Eigenbrod69} and \cite{Bergstrom75}, the present extrapolations employed fits of low-$q$ expansions of the model-independent energy dependence of the form factor with different cutoffs to varying subsets of the low-$q$ data.
In order to match the width of the uncertainty band to the datapoint of Bergstrom \textit{et al.}, the selection of fits had to be limited to a reduced chi-square $\chi^2_\nu$ on the order of $0.1$.
It was found that the width of this band can easily be extended by increasing the cutoff or relaxing the restriction on $\chi^2_\nu$.
The obvious presence of systematic uncertainty in the literature data precludes a comparison to state-of-the art theoretical results \cite{Pastore13, Bacca2014} and calls for a precision measurement directly at the photon point to avoid the extrapolation uncertainty.

We have, therefore, performed an experiment to measure $\Gamma_{\gamma, 0^+_1 \to 1^+_1}$ with the newly developed NRF-based relative SAbs method \cite{Romig15a, Romig15b}.
Compared to the traditional SAbs technique \cite{Metzger1959} used in several previous experiments \cite{Cohen59, Skorka63, Skorka66, *Wahl66, Creten68, Rasmussen69, Saito73}, it utilizes a normalization target (no) in combination with the scattering target of interest (sc) to separate resonant and nonresonant processes \footnote{A detailed derivation of this equation can be found in the supplemental material \cite{SupplementalMaterialNRF}}:
\begin{equation}
\label{experimental_self_absorption}
	R_\mathrm{exp} = 1 - \Bigg\langle \frac{N_\mathrm{no}^\mathrm{nrf}}{N_\mathrm{no}^\mathrm{abs}} \Bigg\rangle \frac{N_\mathrm{sc}^\mathrm{abs}}{N_\mathrm{sc}^\mathrm{nrf}} = R \left( \Gamma_{\gamma, 0^+_1 \to 1^+_1}, T_\mathrm{eff} \right).
\end{equation}
In Eq.~\eqref{experimental_self_absorption}, $N_\mathrm{x}^\mathrm{nrf}$ denotes the number of observed NRF events from a $\gamma$-ray line from material $x$.
The number of events is reduced to $N_\mathrm{x}^\mathrm{abs}$ in a second measurement by the introduction of an absorber target, which consists of the same material as the scatterer of interest, into the incident continuous-energy photon beam.
The reduction of the count rate of the NRF line of interest is due to nonresonant scattering as well as the SAbs induced by the absorber.
Both contributions can be separated in a model-independent way by using the reduction of the count rate in the NRF lines of the normalization target [factor $\langle \nicefrac{N_\mathrm{no}^\mathrm{nrf}}{N_\mathrm{no}^\mathrm{abs}} \rangle$ in Eq.~\eqref{experimental_self_absorption}], which is due to nonresonant effects, only.
In the absence of other decay branches, $R_\mathrm{exp}$ is directly related to $\Gamma_{\gamma, 0^+_1 \to 1^+_1}$ \cite{Metzger1959, Pietralla1995} [see Eq.~\eqref{experimental_self_absorption}], once the thermal motion of the nuclei of interest is taken into account.
It can be treated in terms of an effective temperature $T_\mathrm{eff}$ \cite{Metzger1959, Lamb1939, Pietralla1995} that includes corrections due to condensed-matter effects in the target material (see below).

\begin{figure}[!htb]
\includegraphics[width=0.5\textwidth, trim=10 0 30 30, clip]{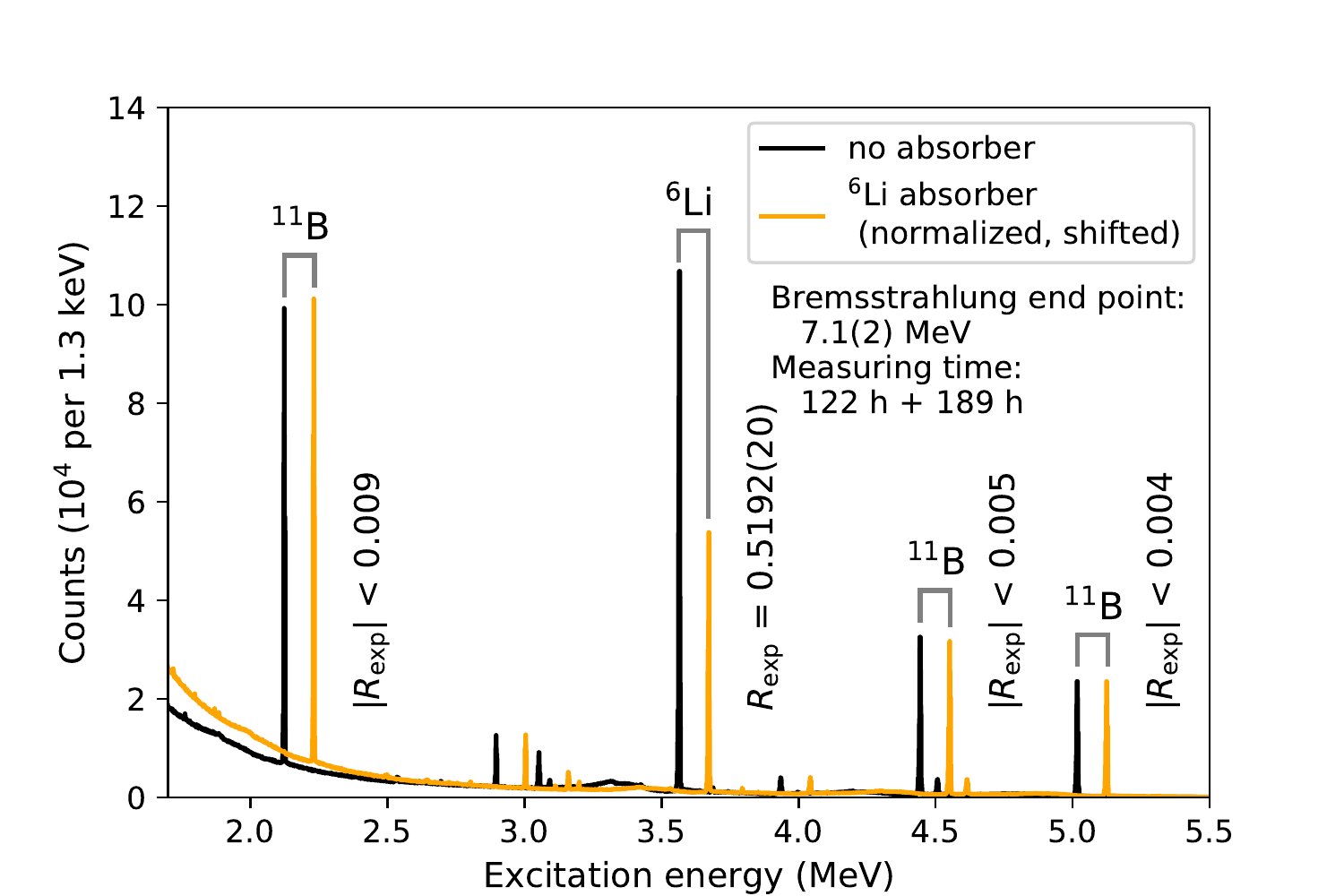}
	\caption{\label{spectrum} Sum spectra of the three detectors from the measurement with (gold) and without (black) the $^6$Li absorber.
	For better visibility, the spectrum with the absorber was shifted by \unit[100]{keV} to higher energies.
	The observed NRF events of three transitions of $^{11}$B were used to normalize the spectrum with the absorber, so that the difference in counts for the $^6$Li transition is due to SAbs only.
	On the right-hand side of the transitions of interest, the (absolute) value of the SAbs ($R_\mathrm{exp}$) is indicated, which is expected to be zero for $^{11}$B.
}
\end{figure}

The experiment was performed at the Darmstadt High-Intensity Photon Setup (DHIPS) \cite{Sonnabend11}, with continuous-energy photon beams generated by bremsstrahlung processes of a \unit[7.1(2)]{MeV} electron beam of the Superconducting Darmstadt Linear Accelerator (S-DALINAC) \cite{Richter96, Pietralla18} on a copper radiator.
A scattering target composed of \unit[5.033(5)]{g} [particle areal density $d_\mathrm{sc,Li} = \unit[0.02773(6)]{\mathrm{b^{-1}}}$, using a target diameter of \unit[20.00(5)]{mm}] of lithium carbonate (Li$_2$CO$_3$) enriched to \unit[95.00(1)]{\%} in $^6$Li, sandwiched between pure boron normalization targets of \unit[2.118(5)]{g} and \unit[2.119(5)]{g} with a \unit[99.52(1)]{\%} $^{11}$B enrichment, was measured for about \unit[122]{h}.
A second, \unit[186]{h}, measurement was carried out with a \unit[9.938(5)]{g} [$d_\mathrm{abs,Li} = \unit[0.05469(10)]{\mathrm{b^{-1}}}$] absorber of the same Li$_2$CO$_3$ material.
Scattered $\gamma$ rays from the target were detected by three high-purity germanium (HPGe) detectors at polar angles of $90^\circ$ (twice) and $130^\circ$ with respect to the beam axis.
To avoid direct scattering of $\gamma$ rays from the absorber target into the detectors, the target was mounted at the entrance of the \unit[1]{m}-long collimation system of DHIPS, which acts as a passive shielding.
The direct scattering into the detectors was found to be negligible by an additional \unit[8]{h} measurement with the absorber target only.
A potential systematic uncertainty due to small-angle scattering of bremsstrahlung $\gamma$ rays inside the collimator, which would then induce excess NRF reactions in the scatterer, was found to be on the order of $\unit[0.33]{\%}$ by {\sc Geant4} \cite{Agostinelli03, Allison06, Allison16} simulations  (i.e., in the anticipated order of magnitude of the uncertainty of $R$) and taken into account by replacing $\langle N^\mathrm{nrf}_\mathrm{no} / N^\mathrm{abs}_\mathrm{no} \rangle$ with  $1.0033 \times \langle N^\mathrm{nrf}_\mathrm{no} / N^\mathrm{abs}_\mathrm{no} \rangle$ in Eq.~\eqref{experimental_self_absorption}.
Summed spectra of all three detectors from the measurements with and without absorber are provided in Fig.~\ref{spectrum}.
Using the known internal $\gamma$-ray transitions of $^{11}$B at 2125, 4445, and \unit[5020]{keV} \cite{Kelley2012}, the measurement with the absorber was normalized to the one without it using the energy-dependent factor $\nicefrac{N_\mathrm{no}^\mathrm{nrf}}{N_\mathrm{no}^\mathrm{abs}}$ in Eq.~\eqref{experimental_self_absorption}.
The normalization factor at the three discrete energies of the $^{11}$B transitions were interpolated by a {\sc Geant4 } simulation of the $\gamma$-ray attenuation, which was in turn validated by an offline measurement with a radioactive $^{56}$Co source.
Including the counting statistics and the correction factor for small-angle scattering, and propagating uncertainties with a Monte-Carlo method \cite{GUM2008}, a value of $R_\mathrm{exp} = 0.5192(20)$ with a relative uncertainty of $\unit[0.39]{\%}$ was obtained.

Li$_2$CO$_3$ was chosen as the target material to reduce systematic uncertainties because pure lithium, used in all previous experiments \cite{Cohen59, Barber60, Booth62, Barber63, Bernheim63, Skorka63, Skorka66, *Wahl66, Creten68, Eigenbrod69, Rasmussen69, Saito73, Bergstrom75}, is highly hygroscopic, which may lead to systematic errors in the determination of the target thickness.
The $T_\mathrm{eff}$ value for Li$_2$CO$_3$ [see Eq.~\eqref{experimental_self_absorption}] was determined from state-of-the-art atomic theory.
First, the phonon density of states (phDOS) of Li$_2$CO$_3$ was obtained from density functional theory (DFT) \cite{Hohenberg1964, Kohn1965}.
Computations of this observable are typically in excellent agreement with experimental data \cite{Gonze1997, Baroni2001}.
The DFT calculations employed the {\sc gpaw} \cite{Mortensen2005, Enkovaara2010} code in a plane-wave basis.
For the exchange-correlation (xc) potential, the local-density (LDA) \cite{Perdew1992} and the generalized-gradient approximation (GGA) \cite{Perdew1996} were tried, which typically slightly under- (LDA) and overestimate (GGA) the crystal binding.
Both xc potentials reproduced the experimental lattice constants $a, b, c$, and $\gamma$ of Li$_2$CO$_3$ \cite{Effenberger1979} with deviations at the $\unit[0.1]{\%}$ level; this can be viewed as a benchmark test.
From the phDOS, a value of $T_\mathrm{eff} = \unit[411(11)]{K}$ was obtained by the procedure described in Ref. \cite{Lamb1939}, which represents the average value and spread of the LDA and GGA solutions.
Using all the aforementioned input \cite{SupplementalMaterialUncertainty}, our experimental value for the $\gamma$-decay width is $\Gamma_{\gamma, 0^+_1 \to 1^+_1} = \unit[8.17_{-0.13}^{+0.14} \left( \mathrm{stat.}\right) _{-0.11}^{+0.10} \left( \mathrm{syst.} \right)]{eV}$, which corresponds to a strength $B(\mathrm{M1}; 0^+ \to 1^+) = \unit[15.61_{-0.25}^{+0.27} \left( \mathrm{stat.}\right) _{-0.21}^{+0.19} \left( \mathrm{syst.} \right)]{\mu_N^2}$.
The \unit[68.3]{\%} coverage interval (CI) is divided into statistical (stat) and systematic (syst) parts, where the latter account for uncertainties in the target dimensions as well as in atomic and condensed-matter contributions\footnote{Since both contributions are uncorrelated and the CIs are almost symmetric, a symmetrized and quadratically summed uncertainty of $\unit[15.61(33)]{\mu_N^2}$ is used in all figures.}.

For the \textit{ab initio} calculations, the importance-truncated no-core shell model (IT-NCSM) \cite{Roth2009, Roth2007} was employed as a state-of-the-art many-body method. 
Within the IT-NCSM, two-nucleon (NN) and three-nucleon (3N) interactions derived within $\chi$EFT were used. 
Four different Hamiltonians (I-IV) were considered, including (I) the Entem-Machleidt (EM) NN interaction at N$^3$LO \cite{EntemMachleidt2003}, complemented with a local 3N interaction (cutoff $\Lambda=500$ MeV, $c_{\text{D}}=0.8$) at N$^2$LO, which is fitted to reproduce the binding energy as well as the $\beta$-decay half-life of $^3$H \cite{Gazit2009, GazitErratum2019}. 
Furthermore, Hamiltonians (II-IV) use the NN interactions by Entem, Machleidt and Nosyk (EMN) at N$^2$LO, N$^3$LO and N$^4$LO with $\Lambda=500$ MeV \cite{EntemMachleidtNosyk2017}, complemented with consistent nonlocal 3N interactions up to N$^2$LO, N$^3$LO and N$^3$LO, respectively.
The NN interactions were only fitted to NN scattering data and the deuteron binding energy, while the 3N interactions were fitted to reproduce the triton binding energy and to optimize the ground-state energy and radius of $^{4}$He, which led to the values $c_{D}=-1$, $c_{D}=2$ and $c_{D}=3$, for the cases II, III and IV, respectively.
The similarity renormalization group (SRG) was employed at the NN and 3N level with a flow parameter of
$\alpha=0.08~\text{fm}^4$ \cite{Roth2011,Roth2014}.

Using an SRG-transformed Hamiltonian requires a consistent SRG transformation of the $M1$ operator.
In previous studies \cite{Calci2016, Shin2017, Parzuchowski2017, Binder2018}, this consistent treatment was neglected, here SRG corrections of the $M1$ operator were included at the two-body level.
In addition to the SRG correction, the NLO 2BC contributions to the $M1$ operator were included as well.
At NLO, these are commonly expressed as a sum of two contributions, the \emph{intrinsic} term and the \emph{Sachs} term~\cite{Pastore09}:
\begin{equation}
\boldsymbol{\mu}^{\text{NLO}}_{[12]}(\mathbf{R}, \mathbf{k}) = \boldsymbol{\mu}^{\text{intrinsic}}_{[12]}(\mathbf{k}) + \boldsymbol{\mu}^{\text{Sachs}}_{[12]}(\mathbf{R}, \mathbf{k})
\end{equation}
with
\begin{eqnarray*}
  \boldsymbol{\mu}^{\text{intrinsic}}_{[12]}(\mathbf{k}) &=& -\frac{i}{2} \nabla_q \times \mathbf{J}(\mathbf{q}, \mathbf{k})|_{\mathbf{q} = 0} \\
  \boldsymbol{\mu}^{\text{Sachs}}_{[12]}(\mathbf{R}, \mathbf{k}) &=& - \frac{i}{2} e(\boldsymbol{\tau}_1 \times \boldsymbol{\tau}_2)_z \, \mathbf{R} \times \nabla_k v(\mathbf{k}) \, .
\end{eqnarray*}
Here, $\boldsymbol{\tau}_i$ are the Pauli matrices, $\mathbf{q}$ the momentum transfer of the photon, $v(\mathbf{k})$ the one-pion exchange potential in momentum representation, and $\mathbf{R}$ the center of mass coordinate of the two nucleons.
The Sachs term only depends on the potential between the two nucleons, whereas the translationally-invariant intrinsic term is given by the spatial part of the two-body current $\mathbf{J}$.
For each interaction, an IT-NCSM calculation was carried out with $N_{\text{max}}$ from 2 to 12 with harmonic-oscillator frequencies $\hbar \Omega=16,20,24$ MeV.
For the resulting value of the magnetic moment and the transition strength, the central value for the highest $N_{\text{max}}$ was used as the nominal result, and the neighboring results as an estimate for the many-body uncertainties.
\begin{figure}
	\includegraphics[width=0.48\textwidth, trim=0 0 42 40,clip]{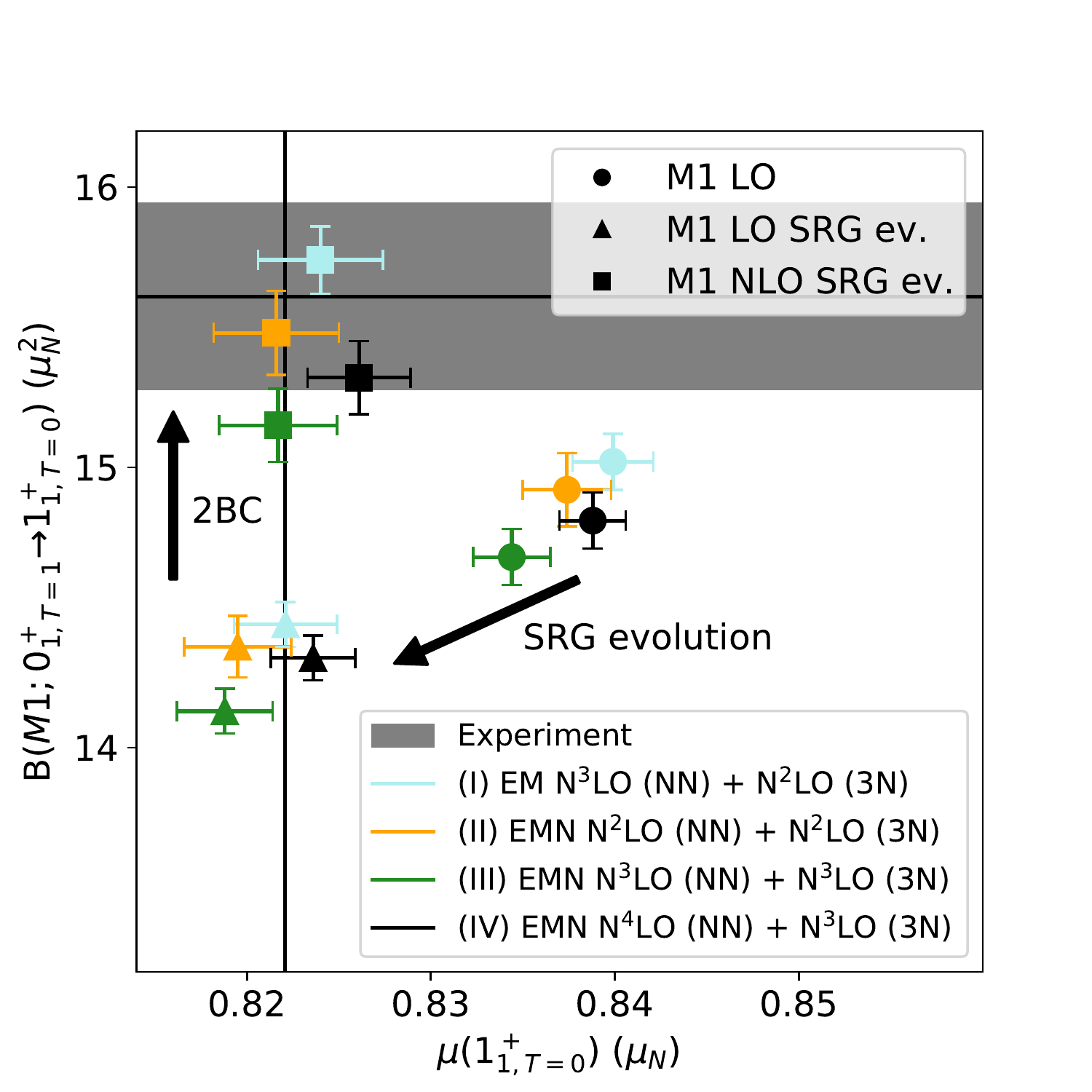}
	\caption{\label{theory_6li}
	Results for $\mathrm{B}(M1; 0^+_{1, T=1} \to 1^+_{1, T=0})$ and $\mu (1^+_{1, T=0})$ from theoretical calculations based on Hamiltonians I-IV (see also Tab.~\ref{tab_theory}).
	As shown in the upper legend, circular markers indicate calculations with the unevolved leading-order (LO) one-body transition operator, triangular markers indicate calculations with the consistently SRG-transformed operator (LO SRG ev.), and quadratic markers indicate the calculations with a consistently SRG-transformed operator including contributions from next-to-leading order 2BC (NLO SRG ev.).
	The labeled arrows illustrate the impact of the two aforementioned improvements.
	Figure~\ref{BM1_overview} shows only the results with the most complete transition operator in the same color code.
	The experimental $68\%$~CI for $B(M1)$ (present work) is indicated by a shaded area, and the most probable values of $B(M1)$ and $\mu$ \cite{Tilley02} by a solid line (the CI of $\mu$ is not visible at this scale).
	}
\end{figure}
The results of the calculations are listed in Tab.~\ref{tab_theory} and displayed in Fig.~\ref{theory_6li} [see also panels (e) and (f) of Fig.~\ref{BM1_overview}], where they are compared to the new experimental constraint of the present work and the magnetic moment $\mu (1^+_{1, T=0}) = \unit[0.82205667(26)]{\mu_N}$ \cite{Tilley02} of the ground state of $^6$Li.
\begin{table}
	\caption{\label{tab_theory}
	Results of the theoretical calculations for $B(M1; 0^+_{1, T=1} \to 1^+_{1, T=0})$ and $\mu(1^+_{1, T=0})$ of $^6$Li.
	These employed four different Hamiltonians (I-IV), which are introduced in the text.
	The calculations are sorted by the type of $M1$ operator, with the same abbreviations as in Fig.~\ref{theory_6li}.
	For comparison, the results of QMC calculations in Refs.~\cite{Marcucci2008, Pastore13} are shown in the second part of the Table.
	The 'standard nuclear physics approach' (SNPA) for the operator in Ref.~\cite{Marcucci2008} was complemented by a $\chi$EFT approach in Ref.~\cite{Pastore13}, while in both cases phenomenological potentials were used.
	'LO' refers to one-body currents and 'Total' to the inclusion of two-body currents.
	}
	\begin{ruledtabular}
		\begin{tabular}{ccccc}
			& I & II & III & IV \\
			\colrule
			\multicolumn{5}{c}{LO} \\
			$\mu$ ($\mu_N$) & 0.8399(22) & 0.8374(24) & 0.8344(21) & 0.8388(18) \\
			$B(M1)$ ($\mu_N^2$) & 15.02(10) & 14.92(13) & 14.68(10) & 14.81(10) \\
			\multicolumn{5}{c}{LO SRG ev.} \\
			$\mu$ ($\mu_N$) & 0.8221(28) & 0.8195(29) & 0.8188(26) & 0.8236(23) \\
			$B(M1)$ ($\mu_N^2$) & 14.44(8) & 14.36(11) & 14.13(8) & 14.32(8) \\
			\multicolumn{5}{c}{NLO SRG ev.} \\
			$\mu$ ($\mu_N$) & 0.8240(34) & 0.8216(34) & 0.8217(32) & 0.8261(28) \\
			$B(M1)$ ($\mu_N^2$) & 15.74(12) & 15.48(15) & 15.15(13) & 15.32(13) \\
			\colrule
			& \multicolumn{2}{c}{\cite{Marcucci2008}} & \multicolumn{2}{c}{\cite{Pastore13}} \\
			\colrule
			\multicolumn{5}{c}{QMC LO} \\
			& \multicolumn{2}{c}{SNPA} & SNPA & $\chi$EFT \\
			$\mu$ ($\mu_N$) & \multicolumn{2}{c}{0.810(1)} & 0.817(1) & 0.817(1) \\
			$B(M1)$ ($\mu_N^2$) & \multicolumn{2}{c}{12.84(11)} & -- & 13.18(4) \\
			\multicolumn{5}{c}{QMC Total} \\
			$\mu$ ($\mu_N$) & \multicolumn{2}{c}{0.800(1)} & 0.807(1) & 0.837(1) \\
			$B(M1)$ ($\mu_N^2$) & \multicolumn{2}{c}{15.00(11)} & -- & 16.07(6) \\
		\end{tabular}
	\end{ruledtabular}
\end{table}

Remarkably, the results of the most complete calculations, including contributions from the 2BC to the $M1$ operator, exhibit an excellent agreement with the new experimental constraints of the present work.
This indicates the importance of 2BC for a correct description of the $^6$Li nucleus.
The residual differences between experiment and theory are probably related to missing higher-order contributions to the $M1$ operator.
The increase of the $B(M1)$ strengh is also found in quantum Monte Carlo (QMC) calculations when 2BC are included \cite{Marcucci2008, Pastore13} (see also Tab.~\ref{tab_theory}).

In contrast to the data points that presently dominate the world average, this measurement was performed directly at the photon point and with controlled systematic uncertainties.
In total, a relative uncertainty of $\unit[2]{\%}$ with balanced contributions by statistics and systematics was achieved.
This translates into an uncertainty of about \unit[2]{attoseconds} for the half-life of the $0^+_1$ state of $^6$Li.
In addition, $\chi$EFT nuclear structure calculations were performed which take into account 2BC at NLO, combined with chiral interactions at various orders, for the first time.
Excellent agreement between experiment and theory was found at a new level of precision in both areas. 

\section{Acknowledgments}
\begin{acknowledgments}
We thank the crew of the S-DALINAC for providing excellent conditions for experimentation.
This work was supported by the Deutsche Forschungsgemeinschaft (DFG) under grants No. SFB 634 (Project ID 5485852), SFB 1044 (204404729), SFB 1245 (279384907), and the Cluster of Excellence PRISMA$^+$ (39083149), by the Bundesministerium f\"ur Bildung und Forschung under grant No. 05P18PKEN9, and by the State of Hesse within the LOEWE research project 'Nuclear Photonics'.
Numerical calculations have been performed on the LICHTENBERG cluster at the computing center of the TU Darmstadt.
This material is based upon work supported by the U.S. Department of Energy, Office of Science, Office of Nuclear Physics, under the FRIB Theory Alliance award DE-SC0013617.
MB, PCR, TB, and UFG acknowledge support by the Helmholtz Graduate School for Hadron and Ion Research of the Helmholtz Association.
RS acknowledges support by the International Max Planck Research School for Precision Tests of Fundamental Symmetries.

\end{acknowledgments}
\appendix
\providecommand{\noopsort}[1]{}\providecommand{\singleletter}[1]{#1}%
\end{document}


\title{Supplemental Material to\\``Role of chiral two-body currents in $^6$Li magnetic properties in light of a new precision measurement with the relative self-absorption technique"}
\date{\today}

\newcommand{\emmi}{\affiliation{ExtreMe Matter Institute EMMI, GSI Helmholtzzentrum f\"ur Schwerionenforschung GmbH, 64289 Darmstadt, Germany}}
\newcommand{\gsi}{\affiliation{GSI Helmholtzzentrum f\"ur Schwerionenforschung GmbH, 64289 Darmstadt, Germany}}
\newcommand{\ikpjguprisma}{\affiliation{Institut f\"ur Kernphysik and PRISMA Cluster of Excellence, Johannes Gutenberg-Universit\"at Mainz, 55128 Mainz, Germany}}
\newcommand{\mainzgsi}{\affiliation{Helmholtz Institute Mainz, GSI Helmholtzzentrum f\"ur Schwerionenforschung GmbH, 64289 Darmstadt, Germany}}
\newcommand{\ikptud}{\affiliation{Institut f\"ur Kernphysik, Technische Universit\"at Darmstadt, 64289 Darmstadt, Germany}
}
\newcommand{\mpi}{\affiliation{Max-Planck-Institut f\"ur Kernphysik, 69117 Heidelberg, Germany}}
\newcommand{\mstud}{\affiliation{Department of Materials Science, Technische Universit\"at Darmstadt, 64287 Darmstadt, Germany}
}
\newcommand{\nsu}{\affiliation{Department of Physics, North Carolina State University, Raleigh, NC 27695, USA}}
\newcommand{\supa}{\affiliation{SUPA, Scottish Universities Physics Alliance, Glasgow, G12 8QQ, UK}}
\newcommand{\triumf}{\affiliation{TRIUMF, 4004 Wesbrook Mall, Vancouver BC, V6T 2A3, Canada}}
\newcommand{\tunl}{\affiliation{Triangle Universities Nuclear Laboratory, Duke University, Durham, NC 27708, USA}}
\newcommand{\ubc}{\affiliation{Department of Physics and Astronomy, University of British Columbia, Vancouver BC, V6T 1Z4, Canada}}
\newcommand{\unc}{\affiliation{Department of Physics and Astronomy, University of North Carolina at Chapel Hill, Chapel Hill, NC 27599, USA}}
\newcommand{\uom}{\affiliation{Department of Physics and Astronomy, University of Manitoba, Winnipeg MB, R3T 2N2, Canada}}
\newcommand{\uws}{\affiliation{School of Engineering, University of the West of Scotland, Paisley, PA1 2BE, UK}}

\author{U.~Friman-Gayer} \email{ufrimangayer@ikp.tu-darmstadt.de} \ikptud \unc \tunl
\author{C.~Romig} \altaffiliation[Present address: ]{Projekttr\"ager DESY, Deutsches Elektronen-Synchrotron, 22607 Hamburg, Germany} \ikptud

\author{T.~H\"uther} \ikptud
\author{K.~Albe} \mstud
\author{S.~Bacca} \ikpjguprisma \mainzgsi
\author{T.~Beck} \ikptud
\author{M.~Berger} \ikptud
\author{J.~Birkhan} \ikptud
\author{K.~Hebeler} \ikptud \emmi
\author{O.~J.~Hernandez} \ubc \ikpjguprisma
\author{J.~Isaak} \ikptud
\author{S.~K\"onig} \ikptud \emmi \nsu
\author{N.~Pietralla} \ikptud
\author{P.~C.~Ries} \ikptud
\author{J.~Rohrer} \mstud
\author{R.~Roth} \ikptud
\author{D.~Savran} \gsi
\author{M.~Scheck} \ikptud \uws \supa
\author{A.~Schwenk} \ikptud \emmi \mpi
\author{R.~Seutin} \mpi \ikptud \emmi
\author{V.~Werner} \ikptud

\maketitle

\tableofcontents

\section{The electron scattering result by Bergstrom \textit{et al.}}

\noindent At present, the most precise experimental value for the reduced transition strength $B(M1;0^+_{1,T=1}~\to~1^+_{1,T=0})$ of $^6$Li has been reported by Bergstrom, Auer and Hicks \cite{Bergstrom75}, using data from their electron scattering experiment, and a previous one by Neuhausen and Hutcheon \cite{Neuhausen71}, obtained at another facility.
Since this measuremet has a significant impact on the currently adopted value for this quantity \cite{Tilley02}, the purpose of this section is to critically assess the uncertainty reported by the authors.
As indicated in the main text, the reduced transition strength is obtained from inelastic electron scattering data by extrapolating the momentum-transfer ($q$) dependence of the squared absolute value of the form factor, $|F(q)|^2$, to the photon point $q_0$. 
Explicitly, for a magnetic transition of multipolarity $\lambda$, the relation is given by \cite{Huby1958}:
\begin{equation}
    \label{form_factor_to_bmlambda}
    |F(q_0)|^2 = \frac{4 \pi \left( 1 + \nicefrac{1}{\lambda} \right) q_0^{2 \lambda}}{Z^2 \left[ \left( 2 \lambda + 1 \right) !! \right]} B(M \lambda; J \to J^\prime).
\end{equation}
In Eq.~\eqref{form_factor_to_bmlambda}, the symbol $Z$ denotes the proton number of the target nucleus. Note that Eq.~\eqref{form_factor_to_bmlambda} contains the reduced transition strength for the excitation from the state with an angular momentum quantum number $J$ to a level $J^\prime$, as opposed to the one for the corresponding decay which is reported in the main text.
The two strengths are related by:
\begin{equation}
    B(M \lambda; J \to J^\prime) = \frac{2J^\prime + 1}{2J + 1} B(M \lambda; J^\prime \to J).
\end{equation}
Bergstrom \textit{et al.} describe their procedure to obtain a quoted uncertainty of $\unit[15.6(4)]{\mu_N^2}$  as follows \cite{Bergstrom75}:
For the extrapolation \textit{``many functions were tried such as polynomials in $q^2$, Helm-type models, as well as more general functions described below, using only low-q data in some cases, and all the available data in others. The fits with the lowest chi-squares per degree of freedom $\chi^2_\nu$ gave $\Gamma_{\gamma,0} (M1) = \unit[8.00-8.23]{eV}$. Our final value is $\Gamma_{\gamma0} (M1) = \unit[8.16 \pm 0.19]{eV}$"}.
Although Ref.~\cite{Bergstrom75} provides all necessary data, these instructions are insufficient to enable an exact reconstruction of the authors' analysis.
To revisit the problem of the form-factor extrapolation, we have made the following conservative assumptions about the selection of data, the fitted model, and the assessment of the goodness of the fit:
\begin{itemize}
    \item The data points of the experiment by Bergstrom \textit{et al.} were taken from Tab.~I in Ref.~\cite{Bergstrom75}.
    The experiment of Neuhausen and Hutcheon was reevaluated in Ref.~\cite{Bergstrom75}, because Bergstrom \cite{Bergstrom752}\footnote{Reference \cite{Bergstrom75}, which was received on May 12, 1975, gives the reference ``\textit{J. C. Bergstrom, Phys. Rev. C, in press}" for the Coulomb correction. Although the article referenced here, Ref.~\cite{Bergstrom752}, was already published on May 1, 1975, i.e., before the submission of the article on $^6$Li, it is highly likely that this is the one referred to in Ref.~\cite{Bergstrom75}.} had derived a new approximative method of correcting for the Coulomb distortion (see, e.g., Ref.~\cite{Theissen1972}) of the electron wave function.
    In addition, a new parameterization of the $^6$Li elastic form factor was available to Bergstrom \textit{et al.} from an experiment by Li \textit{et al.} \cite{Li71}.
    For these reasons, the form-factor data of Neuhausen and Hutcheon were read off from Fig.~1 in Ref.~\cite{Bergstrom75} in our assessment rather than taking them from the original article, Ref.~\cite{Neuhausen71}.
    Table~\ref{tab_ff2} shows the values that were extracted from the Figure as $F(q)$ and converted to $|F(q)|^2$ by Gaussian propagation of uncertainty.
    \begin{table}
        \caption{\label{tab_ff2}Experimental data for the inelastic form factor of the $0^+_1$ state of $^6$Li from an inelastic electron scattering experiment of Neuhausen and Hutcheon \cite{Neuhausen71} ('N').
        The first and the fourth column contain the momentum transfer and the squared absolute value of the form factor as given in the original publication \cite{Neuhausen71}, respectively.
        For the second and third column, the form factor $F(q)$ was read off from Fig.~2 in the article by Bergstrom \textit{et al.} \cite{Bergstrom75} ('B') and converted to $|F(q)|^2$.
        The values in the third and fourth column are not identical due to the use of different corrections for the Coulomb distortion.
        }
        \begin{ruledtabular}
            \begin{tabular}{crrr}
            $q$ (fm$^{-1}$) (N) & $F(q)$ (B) & $|F(q)|^2$ (B) & $|F(q)|^2$ (N) \\
            \colrule
            0.754 & $3.56(9) \times 10^{-2}$ & $1.27(6) \times 10^{-3}$ & $1.31(6) \times 10^{-3}$ \\
            0.859 & $3.09(8) \times 10^{-2}$ & $9.6(5) \times 10^{-4}$ & $1.00(5) \times 10^{-3}$ \\
            0.954 & $2.33(8) \times 10^{-3}$ & $5.4(4) \times 10^{-4}$ & $5.7(4) \times 10^{-4}$ \\
            1.047 & $1.87(6) \times 10^{-2}$ & $3.49(26) \times 10^{-4}$ & $3.7(3) \times 10^{-4}$ \\
            1.182 & $8.8(14) \times 10^{-3}$ & $7.8(26) \times 10^{-5}$ & $8.6(30) \times 10^{-5}$ \\
            1.226 & $7.6(12) \times 10^{-3}$ & $5.7(18) \times 10^{-5}$ & $6.3(19) \times 10^{-5}$ \\
            1.525 & $5.5(9) \times 10^{-3}$ & $3.0(10) \times 10^{-5}$ & $3.4(12) \times 10^{-5}$ \\
            1.680 & $8.7(7) \times 10^{-3}$ & $7.6(12) \times 10^{-5}$ & $8.8(18) \times 10^{-5}$ \\
            1.747 & $8.8(11) \times 10^{-3}$ & $7.7(19) \times 10^{-5}$ & $9.0(22) \times 10^{-5}$ \\

            \end{tabular}
        \end{ruledtabular}
    \end{table}
    The corrections by Bergstom \textit{et al.} change the values of $|F(q)|^2$ by about \unit[3]{\%} at low $q$ values up to \unit[14]{\%} at high $q$ values.
    The authors of Ref. \cite{Bergstrom75} did not use the data points of Neuhausen and Hutcheon at $q = \unit[0.696]{fm^{-1}}$ and $\unit[1.363]{fm^{-1}}$ for the fit, so they are also neglected here.
    In total, $n_{d, \mathrm{max}}= 27$ data points are available for the fit.
    Bergstrom \textit{et al.} report that, sometimes, they only used a subset of $n_d < n_{d, \mathrm{max}}$ data points.
    In principle, the instructions given in Ref.~\cite{Bergstrom75} do not place any restrictions on which  $n_d$ data points to choose, but we assume that $n_d$ means the $n_d$ data points with the lowest values of the momentum transfer.
    \item To study the impact of the form-factor model, the `model-independent' approach of Chernykh \textit{et al.} \cite{Chernykh2010, DAlessio2020} was used, which is based on a low-$q$ expansion of transition form factors:
    \begin{align}
        \label{form_factor_model}
        F(q, n_\mathrm{max}, b, \lbrace c_n \rbrace) = & \exp \left[-\frac{1}{2} \left( bq \right)^2 \right] \\
        & \times \sum_{n=1}^{n_\mathrm{max}} c_n \left( bq \right)^{2n}. \nonumber
    \end{align}
    Equation~\eqref{form_factor_model} contains $n_\mathrm{max}$ expansion parameters $c_n$ and another parameter $b$, i.e., $n_\mathrm{max}+1$ parameters in total.
    This model is supposed to be a compromise between the polynomials, unspecified `Helm-type models' and `more general functions' described by Bergstrom \textit{et al.}, and it allows for a systematic increase of the complexity by increasing the maximum index of the sum.
    \item To assess the goodness of their fits, Bergstrom \textit{et al.} \cite{Bergstrom75} calculated the chi-square statistic per degree of freedom ($\chi^2_\nu$)~\footnote{Although there are more sophisticated methods to determine the goodness of a fit, it was decided to use the same criterion as in the original publication. The relative uncertainties of the form-factor data are low, so that the $\chi^2_\nu$ statistic is expected to be a reasonable approximation.}, which is defined in this case as (for the general definition, see, e.g., the statistics section in the 'Review of Particle Physics' \cite{Zyla2020}):
    \begin{align}
        \label{chi_square}
        &\chi^2_\nu = \frac{1}{n_d - n_\mathrm{max} - 1} \\
        & \times \sum_{i = 1}^{n_d} \frac{\lbrace | F_i \left( q_i \right) |^2 - | F \left( q_i, n_\mathrm{max}, b, \lbrace c_n \rbrace \right) |^2 \rbrace^2}{\Delta \left[ | F_i \left( q_i \right) |^2 \right]}. \nonumber
    \end{align}

    The sum in Eq.~\eqref{chi_square} runs over all data points considered in the fit.
    The symbol $\Delta \left[ | F_i \left( q_i \right) |^2 \right]$ denotes the uncertainty of the $i$-th data point.

    Bergstrom \textit{et al.} report an interval of possible transition widths from an arbitrary selection of fits with the lowest $\chi^2_\nu$ values.
    Although it is not stated how their `final value' was deduced from the fits, this result is usually interpreted \cite{AjzenbergSelove79, AjzenbergSelove88, Tilley02} as a \unit[68.3]{\%} coverage interval (the range $\mu \pm 1\sigma$ for a normal distribution with mean value $\mu$ and standard deviation $\sigma$).

    Here, the goodness of the fits was assessed by their $p$ value (see, e.g., Ref.~\cite{Zyla2020}).
    In accordance with Eq.~\eqref{chi_square}, a $\chi^2$ distribution was chosen as a statistic.
    This means that a reasonable fit should yield a $\chi^2_\nu$ value on the order of 1. 
    A value of $p=0.317$ was chosen in analogy to a \unit[68.3]{\%} coverage interval.
    Note that there are also procedures for obtaining the statistical coverage interval of a single fit \cite{Zyla2020}, but the deviations between different models were so large that this contribution was found to be negligible.
\end{itemize}
In order to get a comprehensive overview of the model- and data-set dependence, the parameters $n_d$ and $n_\mathrm{max}$ of the analysis were varied within the following limits:
\begin{equation}
    \label{limits_n_d}
    3 \leq n_d \leq n_{d, \mathrm{max}} = 27,
\end{equation}
\begin{equation}
    \label{limits_n_max}
    1 \leq n_\mathrm{max} < n_d - 1.
\end{equation}
Equation~\eqref{limits_n_max} ensures that the number of degrees of freedom is positive and nonzero.
Since any nontrivial expansion of the form factor according to Eq.~\eqref{form_factor_model} has at least two parameters, at least three data points were always used.

\begin{figure}
    \includegraphics[width=0.5\textwidth, trim=40 60 40 70, clip]{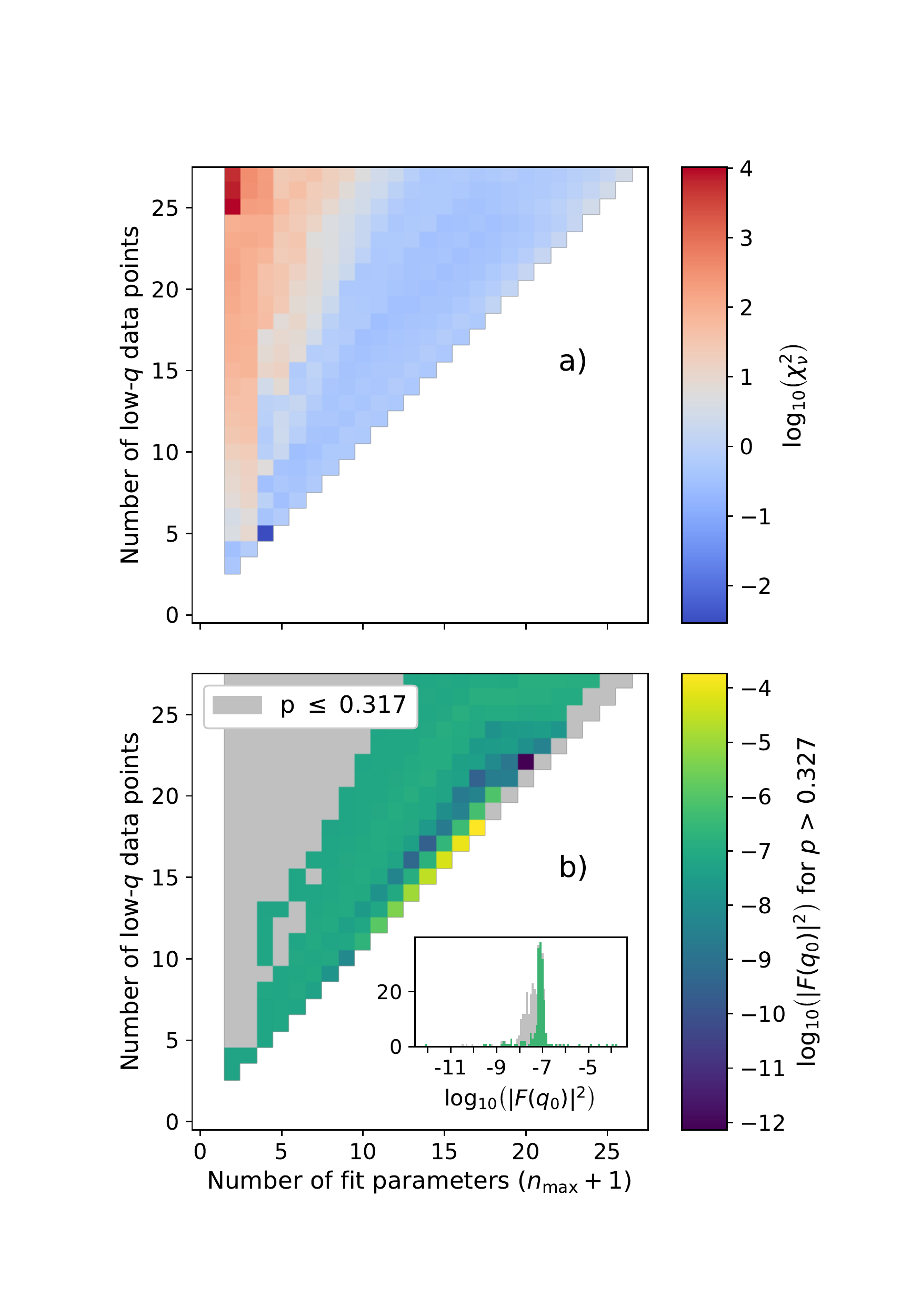}
    \caption{
        \label{form_factor_fit_overview} Results of a series of fits to the $(e,e')$ inelastic form factor of Bergstrom \textit{et al.} \cite{Bergstrom75, Neuhausen71}.
        All fits employed a model-indenpendent low-$q$ expansion of the form factor [Eq.~\eqref{form_factor_model}].
        The cutoff parameter $n_\mathrm{max}$ of this model, as well as the number of data points considered for the fit, were systematically varied [Eqs.~\eqref{limits_n_d} and \eqref{limits_n_max}].
        (a): The chi square per degree of freedom $\chi^2_\nu$ for each fit [Eq.~\eqref{chi_square}, logarithmic scale], which was used as an estimate of the goodness of the fits.
        (b): The values of the form factor at the photon point $|F(q_0)|^2$, extracted from extrapolations of the fitted models (logarithmic scale). 
        Results from fits with a $p$ value lower or equal to 0.317 are shown in a uniform grey color.
        The colored pixels indicate all fits with a $p$ value larger than 0.317.
        The inset shows a histogram of the results for fits with $p > 0.317$ (green), and a histogram of all results [grey, i.e. including the grey-shaded pixels in b)].
    }
\end{figure}

\noindent The results of the 325 individual fits of the present analysis are summarized in Fig.~\ref{form_factor_fit_overview}.
It can be seen that, even when fits are selected by reasonable $p$ values, the extrapolated values of the form factor at the photon point $q_0$ vary by orders of magnitude.
The median and the shortest \unit[68.3]{\%} coverage interval of the extrapolated values, obtained from the fits with $p > 0.317$, are $|F(q_0)|^2 = 8^{+5}_{-2} \times 10^{-8}$ (Note that the relative uncertainty of $|F(q_0)|^2$ is approximately the same as the one for the transition strength or the level width due to the negligible uncertainty in the excitation energy \cite{Tilley02}.).
As already mentioned in the main text, we note that Bergstrom \textit{et al.} \cite{Bergstrom75} claim to have made a similar assessment of the goodness of their fits.
However, a coverage interval close to the one reported by them can only be obtained by restricting the accepted $\chi^2_\nu$ to low values, and does not appear to be based on a quantitative statistical argument.
As expected, the most extreme deviations from this shortest coverage interval are found when the number of data points is close to the number of fit parameters due to the phenomenon of 'overfitting'.
In summary, our systematic reanalysis of the experimental data of Bergstrom \textit{et al.} indicates that the information given in Ref.~\cite{Bergstrom75} is not sufficient to replicate their result.
The authors may have used more restrictive assumptions or systematic errors may have been neglected.
Considering present-day demands on the accuracy of experimental data and on transparency in uncertainty assignments, we consider this situation insufficient as a firm basis of the strength of this gamma-ray transition.

\noindent We would also like to point out that the publication of the second-most precise electron scattering experiment by Eigenbrod \cite{Eigenbrod69} gives a detailed account of their statistical and systematic uncertainties.
However, the model dependence of the extrapolation, which was shown to have a large impact here, was not investigated.
\section{Self-absorption experiment}
\noindent In this section, details about the data analysis in the relative self absorption (SAbs) \cite{Romig15a, Romig15b} experiment are given with a focus on the different contributions to the reported uncertainty.
A preliminary analysis has already been published in the Ph.D. thesis of C. Romig \cite{Romig15b}.
The first two parts of this section connect experimental observables (count rates) to the property of interest, $\Gamma_\gamma$.
This is followed by a presentation of the actual data analysis.
\subsection{Nuclear resonance fluorescence}
\noindent The formalism of nuclear resonance fluorescence (NRF) \cite{KneisslPitzZilges1996, Metzger1959} is closely related to the quantum excitation of an isolated resonance \cite{BreitWigner1936, Lamb1939}.
Consequently, a measure for the probability for the absorption of a photon with an energy $E$ by a nucleus at rest in its ground state with spin $J_0$ is given by the Breit-Wigner cross section:
\begin{equation}
    \label{breit_wigner_cross_section}
    \sigma_{0 \to x} \left( E \right) = \frac{\pi}{2} \left( \frac{\hbar c}{E_x} \right)^2 \frac{2J_x + 1}{2J_0 + 1} \frac{\Gamma_{0 \to x} \Gamma_x}{\left( E - E_x \right)^2 + \nicefrac{\Gamma_x^2}{4}}.
\end{equation}
In Eq.~\ref{breit_wigner_cross_section}, $E_x$ and $J_x$ denote the energy with respect to the ground state (`0') and the spin, respectively, of the isolated excited state (`x') that is populated by the reaction.
The quantities $\Gamma_x$ and $\Gamma_{0 \to x}$ denote the total width of the excited state and the partial width for the excitation of this state from the ground state.
In the case of the $0^+_1$ state of $^6$Li, it is known \cite{Robertson84} that the gamma-decay width to the ground state differs from the total width by less than a fraction of $10^{-7}$.
Therefore, $\Gamma_x = \Gamma_{0 \to x} \equiv \Gamma_\gamma$ is used here.

\noindent For an ensemble of capturing nuclei with a distribution $p(v_z)$ of the velocity component along the $z$ axis of the photon beam, Eq.~\ref{breit_wigner_cross_section} is modified to take into account the Doppler shift $E \to E \left( v_z \right)$ of the gamma-ray energy in the rest frame of the nucleus:
\begin{equation}
    \label{doppler_broadened_cross_section}
    \sigma_{\mathrm{D}, 0 \to x} \left( E \right) = \int_{-\infty}^{\infty} \sigma_{0 \to x} \left[ E (v_z) \right] p \left( v_z \right) \mathrm{d} v_z.
\end{equation}
Equation~\eqref{doppler_broadened_cross_section} represents a semiclassical procedure of taking into account the presence of the capturing nuclei in a system at finite temperature \cite{Metzger1959, Romig15b}.
Quantum mechanical calculations in real \cite{vanHove1954, SingwiSjoelander1960} and momentum space \cite{Lamb1939} show that the binding in an isotropic atomic crystal can be corrected for by the choice
\begin{equation}
    \label{velocity_distribution}
    p(v_z) = \sqrt{\frac{M}{2 \pi k_B T_\mathrm{eff}}} \exp \left( - \frac{M v_z^2}{2 k_B T_\mathrm{eff}} \right),
\end{equation}
i.e., a Maxwell-Boltzmann distribution for nuclei with masses $M$ at an effective temperature $T_\mathrm{eff}$.
All information about the dynamics in the crystal lattice at a finite temperature $T$ is contained in the product of the Boltzmann constant $k_B$ and the effective temperature, which can be interpreted as the average phonon energy of the system \cite{Lamb1939}.
In the present analysis, the phonon density of states (phDOS) for the Li$_2$CO$_3$ crystal was obtained from a state-of-the-art DFT calculation as described in the main text.

\noindent It should be noted that the energy-integrated cross section for the absorption, $I_{0 \to x}$, is independent of the velocity distribution of the nuclei, in particular:
\begin{equation}
    \label{integrated_cross_section}
    I_{0 \to x} = \int_0^\infty \sigma_{0 \to x} \left( E \right) \mathrm{d}E = \int_0^\infty \sigma_{D, 0 \to x} \left( E \right) \mathrm{d}E.
\end{equation}
In the thin-target approximation (see below), the experimental count rate is simply proportional to $I_{0 \to x}$.
For a sufficiently thick target, however, a beam of photons that passes through matter will undergo a significant attenuation.
One contribution to this attenuation, the self-absorption effect, is sensitive to the transition width $\Gamma_\gamma$.
\subsection{Nuclear resonance fluorescence on a thick composite target}
\noindent In this section, the relation between the number of detected NRF reactions $N_{0 \to x \to y, t} \left( ^A_Z\mathrm{X}\right)$ corresponding to an excitation of the state $x$ of an isotope $^A_Z\mathrm{X}$ and the subsequent decay to a state $y$ in the target layer $t$, and the quantity of interest, $\Gamma_{0 \to x}$, is derived for a target consisting of multiple layers with the same geometrical cross section.
From upstream to downstream, along the direction of propagation of the beam, the target layers are the $^6$Li absorber ($t = \mathrm{a}$), the $^{11}$B normalization target upstream of the $^6$Li scatterer ($t = \mathrm{u}$), the $^6$Li scatterer ($t = \mathrm{s}$), and the $^{11}$B normalization target downstream of the $^6$Li scatterer ($t = \mathrm{d}$)\footnote{In principle, the beam also traverses an about \unit[1.5]{m}-long (see below) volume of air, but since the impact on all measurements is the same, it is neglected here.}.
Although they are identified by the respective isotope of interest, each single layer may contain multiple chemical elements X, which in turn may encompass several isotopes $^A_Z$X.

\noindent In general, $N_{0 \to x \to y, t}\left( ^A_Z\mathrm{X}\right)$ can be obtained from a three-dimensional integral over the product of the energy-differential time-integrated position-dependent photon flux $\Phi \left( E, \mathbf{x} \right)$ (the number of photons per unit area and energy interval, simply called `photon flux' in the following), the energy- and direction-dependent $\gamma$-ray detection efficiency $\epsilon \left( E, \mathbf{x} \right)$, the $\gamma$-ray angular correlation $W_{0 \to x \to y} \left( \mathbf{x} \right)$, the energy-dependent absorption cross section, multiplied by the branching ratio for the subsequent decay to the level $y$, $\sigma_{\mathrm{D}, 0 \to x} \left( E \right) \nicefrac{\Gamma_{x \to y}}{\Gamma_x}$, and the position-dependent particle density (nuclei per unit volume) of an isotope, $n_t \left(^A_Z\mathrm{X}, \mathbf{x} \right)$:
\begin{widetext}
\begin{align}
    \label{nrf_reactions_general}
    N_{0 \to x \to y, t} \left( ^A_Z \mathrm{X} \right) = & \frac{\Gamma_{x \to y}}{\Gamma_x} \iiint_\mathrm{V_t} \int_0^\infty \Phi \left( E, \mathbf{x}_t \right) \sigma_{\mathrm{D}, 0 \to x} \left( E \right) \iint_{S_d} \epsilon \left( E, \mathbf{x}_d - \mathbf{x}_t \right) W_{0 \to x \to y} \left( \mathbf{x}_d - \mathbf{x}_t \right) n_t \left( ^A_Z \mathrm{X}, \mathbf{x}_t \right) \mathrm{d} \mathbf{x}_d  \mathrm{d} E \mathrm{d} \mathbf{x}_t
\end{align}
\end{widetext}
In Eq.~\eqref{nrf_reactions_general}, the two- and three-dimensional integrals are assumed to be over all points $\mathbf{x}_d$ on the detector surface area ($S_d$) and all points $\mathbf{x}_t$ in the target layer volume ($V_t$), respectively.
At the precision level of the present experiment, i.e., a relative uncertainty of $R_\mathrm{exp}$ of ${3.9 \times 10^{-3}}$, the position- and energy dependence of many quantities in the equation is negligible.
Furthermore, many of the complicated factors need not be considered explicitly, since $R_\mathrm{exp}$ is a ratio of expressions like Eq.~\eqref{nrf_reactions_general}.

\noindent An important simplification is the assumption that the propagation of the beam through the targets can be treated as a one-dimensional problem on the $z$ axis.
First of all, this requires that the mass distribution inside the targets is homogenous, i.e.:
\begin{equation}
    \label{approximation_homogenous_target}
    n_t \left( ^A_Z\mathrm{X}, \mathbf{x} \right) \approx n_t \left( ^A_Z\mathrm{X} \right).
\end{equation}
Since the finely powdered lithium carbonate and boron material was pressed into plastic tubes with due care \cite{Romig15b}, it is assumed that Eq.~\eqref{approximation_homogenous_target} was fulfilled at the precision level of the present experiment \footnote{Nevertheless, the uncertainty of the absolute value of $n_t$ is dominated by the relative uncertainty of the inner geometrical cross section area of the plastic tubes, which was on the order of $1 \times 10^{-2}$. 
This was taken into account in the calculation of the total uncertainty (see below)}.

\noindent Secondly, the divergence of the photon beam has to be negligible:
\begin{equation}
    \Phi \left( E, \mathbf{x}_t \right) \approx \Phi \left( E, z_t \right).
\end{equation}
This condition was fulfilled due to the characteristics of the experimental setup:
The photon beam was created by the electron beam of the S-DALINAC impinging on a copper radiator target.
Compared to all other dimensions in the setup, the spot size of the electron beam is small \cite{Pietralla18}.
In addition, the stopping power of the copper target for the electron energy of interest is large \cite{Berger2017}, so that the photon beam can be considered to originate from a single point along the optical axis.
The radiator is located about \unit[30]{cm} upstream from the entrance of the collimation system of the Darmstadt High-Intensity Photon Setup (DHIPS) \cite{Sonnabend11}, where the absorber target is mounted.
The scattering targets, in turn, are located more than \unit[1]{m} downstream from the absorber, making the distance from the photon source more than \unit[1.5]{m}.
Although the absorber covers a much larger solid angle than the scatterer, only the photon paths that could result in an NRF reaction on the scatterer have an impact on the experimental count rate.
To a good approximation (see also the correction for small-angle scattering in the collimator below), this geometry leads to an effective solid angle of about \unit[0.14]{mrad}.
When the solid angle is varied by this amount, the simple but sufficiently accurate (see, e.g., Ref.~\cite{Schwengner2005}) analytical expression by Schiff \cite{Schiff1951} predicts a relative variation of the photon beam intensity distribution by less than $10^{-4}$.
Therefore, Eq.~\eqref{nrf_reactions_general} can be simplified to:
\begin{widetext}
    \begin{align}
    \label{nrf_reactions_z_axis}
    N_{0 \to x \to y, t} \left( ^A_Z \mathrm{X} \right) = & \frac{\Gamma_{x \to y}}{\Gamma_x} n_t \left( ^A_Z \mathrm{X} \right) \int_{z_t} \int_0^\infty \Phi \left( E, z_t \right) \sigma_{\mathrm{D}, 0 \to x} \left( E \right) \iint_\mathrm{A_t} \iint_{S_d} \epsilon \left( E, \mathbf{x}_d - \mathbf{x}_t \right) \\
    &\phantom{\frac{\Gamma_{x \to y}}{\Gamma_x} n_t \left( ^A_Z \mathrm{X} \right) \int_{z_t}} \times W_{0 \to x \to y} \left( \mathbf{x}_d - \mathbf{x}_t \right) \mathrm{d} \mathbf{x}_d \mathrm{d} x_t \mathrm{d} y_t \mathrm{d} E \mathrm{d} z_t \nonumber.
\end{align}
\end{widetext}
Here, the symbol $A_t$ denotes the cross-section area of the target, and the volume integral has been split up into integrals over the three coordinates $x$, $y$, and $z$.

\noindent In the one-dimensional coordinate system, the dependence of the photon flux on the energy and penetration depth $z$ into the target $t$ is governed by the differential equation:
\begin{widetext}
\begin{equation}
    \label{photon_flux_differential_equation}
    \frac{\mathrm{d} \Phi}{\mathrm{d} z} \left( E, z \right) = -\left\{ \sum_{^A_Z \mathrm{X} \in t} \left[ \sigma_{\mathrm{nr}, t} \left( E; ^A_Z \mathrm{X} \right) + \sum_{x \in ^A_Z \mathrm{X}} \sigma_{0 \to x, t} \left( E; ^A_Z \mathrm{X} \right) \right] n_t \left( ^A_Z \mathrm{X} \right) \right\} \Phi \left( E, z \right).
\end{equation}
\end{widetext}
Equation~\eqref{photon_flux_differential_equation} contains sums over all isotopes in a target ($^A_Z \mathrm{X} \in t$) and all excited states of a single isotope ($x \in ^A_Z\mathrm{X}$).
The decay of the photon flux is due to nonresonant processes such as Compton scattering and electron-positron pair production, whose cross section is summarized as $\sigma_\mathrm{nr}$ here, and to the resonant absorption described in the previous section.

\noindent The Doppler-broadened resonance with the largest full width at half maximum (FWHM) of all relevant isotopes in the present experiment [$\mathrm{FWHM}_{3/2^-_2} \left( ^{11}\mathrm{B} \right) = \unit[26.5]{eV}$\footnote{Estimated using the Debye approximation \cite{Lamb1939, Metzger1959} with the Debye temperature of natural boron at room temperature, $T_\mathrm{D} = \unit[1362]{K}$, from Ref.~\cite{Ho1972}.}] is the $3/2^-_2$ state of $^{11}$B at \unit[5020]{keV} \cite{Kelley2012}.
On the other hand, all resonance energies are separated by hundreds of keV \cite{Tilley02, Kelley2012, Kelley2017, Tilley1993}.
Therefore, each resonance can be considered as being isolated within a sufficiently small energy range $\left[ E_x - \Delta E_x, E_x + \Delta E_x \right]$ around the resonance energy.
Here, $\Delta E_x = 32 \times \mathrm{FWHM}_x \approx \unit[1]{keV} $ was chosen, which corresponds to a confidence interval (CI) of more than \unit[99]{\%} of a Breit-Wigner cross section with the given FWHM.

\noindent Furthermore, since it is difficult to distinguish nonresonant contributions by different elements, an effective nonresonant attenuation coefficient is introduced\footnote{The symbol $\mu$ is a deliberate reference to the x-ray mass attenuation coefficients of Hubbell and Seltzer \cite{HubbellSeltzer2004} which are often used to correct for the nonresonant attenuation of the photon beam.}:
\begin{equation}
    \label{effective_nonresonant_cross_section}
    \mu_t \left( E \right) \equiv \sum_{^A_Z \mathrm{X} \in t} \sigma_{\mathrm{nr}, t} \left( E; ^A_Z \mathrm{X} \right) n_t \left( ^A_Z \mathrm{X} \right).
\end{equation}
In the excitation energy range of the present experiment, i.e., a few MeV, the cross sections for all nonresonant processes vary smoothly with energy.
For example, within the energy range $\Delta E_x$ chosen above, one finds that
\begin{align}
    \label{constant_mass_attenuation_boron}
    &\left| \frac{\mu_{u,d} \left( E_{3/2^-_2} + \Delta E_{3/2^-_2} \right) - \mu_{u,d} \left( E_{3/2^-_2} - \Delta E_{3/2^-_2} \right)}{\mu_{u,d} \left( E_{3/2^-_2} \right)} \right| \\
    &\approx 2 \times 10^{-4} \nonumber
\end{align}
from a linear interpolation of the tabulated data in Ref.~\cite{HubbellSeltzer2004}.
Therefore, the additional approximation can be made that the attenuation coefficient is constant within the range of the resonance:
\begin{equation}
    \label{constant_mass_attenuation}
    \mu_t \left(E \right) \approx \mu_t \left( E_x \right)
\end{equation}
\begin{equation*}
    \mathrm{for}~E~\mathrm{in}~\left[E_x - \Delta E_x, E_x + \Delta E_x \right].
\end{equation*}
Since the detection efficiency $\epsilon$, i.e., the probability of a photon being absorbed in the detector crystal, is closely related to the attenuation coefficient, the approximation
\begin{equation}
    \label{constant_efficiency}
    \epsilon \left(E \right) \approx \epsilon \left( E_x \right)
\end{equation}
\begin{equation*}
    \mathrm{for}~E~\mathrm{in}~\left[E_x - \Delta E_x, E_x + \Delta E_x \right].
\end{equation*}
can be made in analogy to Eq.~\eqref{constant_mass_attenuation} \footnote{Compare also a typical energy dependence of the efficiency depicted in Ref.~\cite{Romig15b}.}.

\noindent With these simplifications, Eq.~\eqref{photon_flux_differential_equation} becomes:
\begin{align}
    \label{photon_flux_differential_equation_simplified}
    &\frac{\mathrm{d} \Phi}{\mathrm{d} z} \left( E, z \right) \\
    & =
    -\left[ \mu_{t} \left( E_x \right) + \sigma_{0 \to x, t} \left( E; ^A_Z \mathrm{X} \right) n_t \left( ^A_Z \mathrm{X} \right) \right]\Phi \left( E, z \right). \nonumber
\end{align}
In the following, let the origin of the $z$ axis be the most upstream part of the target and denote the initial photon flux as $\Phi \left( E, 0 \right)$.
In addition, the points $z_{0,t}$ and $z_{1,t}$ denote the start- and end point of a target layer.
By using the approximate energy dependence of the photon beam intensity distribution of Schiff \cite{Schiff1951}, one obtains results such as
\begin{equation}
    \label{constant_photon_flux_lithium}
    \left| \frac{\Phi \left( E_{0^+_1} + \Delta E_{0^+_1} \right) - \Phi \left( E_{0^+_1} - \Delta E_{0^+_1} \right)}{\Phi \left( E_{0^+_1} \right) } \right| \approx 8 \times 10^{-4}.
\end{equation}
Therefore, just like the cross section for nonresonant scattering, the initial photon flux can be considered to be constant within the range of a single resonance:
\begin{equation}
    \label{constant_photon_flux}
    \Phi \left( E, 0 \right) \approx \Phi \left( E_x, 0 \right)
\end{equation}
\begin{equation*}
    \mathrm{for}~E~\mathrm{in}~\left[E_x - \Delta E_x, E_x + \Delta E_x \right].
\end{equation*}
Before the solution of Eq.~\eqref{photon_flux_differential_equation_simplified} is presented, it should be noted that, in general, Eq.~\eqref{photon_flux_differential_equation} also contains a term that replenishes the photon flux at an energy $E$ as the beam passes through the setup.
This is due to the `notch-refilling' effect \cite{Pruet2006, Vavrek2019}, which is dominated, at the energies of interest in the present experiment, by small-angle Compton scattering of photons with energies larger than $E$.
The scattering may occur either in the targets themselves or in the environment.
Pruet \textit{et al.} \cite{Pruet2006} have simulated the impact of this effect in the former case for different values of the `optical density' 
\begin{equation}
    \label{optical_density}
    \tau = \mu_{x,t} \left( z_{1,t} - z_{0, t} \right),
\end{equation}
using the particle simulation toolkit MCNP5 \cite{MCNP2003}.
In the present experiment, even the upper limit for the optical density of the Li$_2$CO$_3$ absorber of $\tau_\mathrm{a} = 0.11$, estimated using the x-ray mass attenuation coefficient for oxygen of Hubbell and Seltzer \cite{HubbellSeltzer2004}, is low enough for the notch-refilling effect to be on the order of $< 10^{-3}$ \footnote{
Compare to the optical-depth dependence of the `notch refilling fraction' for `$d=\unit[1]{cm}$, Brem. source' in Fig.~4 of Ref.~\cite{Pruet2006}. 
Note that the authors have assumed a constant power spectrum of the bremsstrahlung beam. 
Actually, the photon beam intensity distribution decreases approximately exponentially over large energy intervals \cite{Schiff1951}, which would make the notch-refilling effect even weaker.}.
On the other hand, the notch refilling by small-angle scattering of beam photons in the about \unit[1]{m}-long collimator of DHIPS has a significant impact on the number of detected NRF events when the absorber target is present.
It amounts to \unit[0.33]{\%} as mentioned in the main text.
The artificial increase of the count rate was corrected for with a multiplicative factor obtained from Geant4 \cite{Agostinelli03, Allison06, Allison16} simulations \cite{Romig15b}.
Simulations of this kind at energies of few MeV typically exhibit a relative deviation from experimental data by less than \unit[20]{\%} \cite{Vavrek2019, Mayer2020}, even for intricate geometries with dozens of interactions per photon.

\noindent In this case, the solution of Eq.~\eqref{photon_flux_differential_equation_simplified} inside a target layer $t$ can be defined recursively as:
\begin{widetext}
\begin{equation}
    \label{photon_flux}
    \Phi(E, z) = \Phi(E, z_{0, t}) \exp \left\{ -\left[ \mu_{t} \left( E_x \right) + \sigma_{0 \to x, t} \left( E; ^A_Z\mathrm{X} \right) n_t \left( ^A_Z \mathrm{X} \right) \right] \left( z - z_{0, t} \right) \right\}
\end{equation}
\begin{equation*}
    \mathrm{for}~z~\mathrm{in}~\left[ z_{0,t}, z_{1,t} \right].
\end{equation*}
\end{widetext}
Using Eqs.~\eqref{constant_mass_attenuation}, \eqref{constant_efficiency}, \eqref{constant_photon_flux}, and \eqref{photon_flux} in Eq.~\eqref{nrf_reactions_z_axis} yields:
\begin{widetext}
\begin{align}
    \label{nrf_reactions_energy_independence}
    N_{0 \to x \to y, t} \left( ^A_Z \mathrm{X} \right) = \frac{\Gamma_{x \to y}}{\Gamma_x} n_t \left( ^A_Z \mathrm{X} \right) A_t \Phi \left( E_x, 0 \right) \int_{E_x - \Delta E_x }^{E_x + \Delta E_x } \int_{z_{0, t}}^{z_{1, t}} \alpha_{0 \to x, t} \left( E, z_t, ^A_Z\mathrm{X} \right) \langle \epsilon W \rangle_{0 \to x \to y, t} \left( E_x, z_t \right) \mathrm{d} z_t \mathrm{d} E.
\end{align}
\end{widetext}
Equation~\eqref{nrf_reactions_energy_independence} contains two abbreviations inside the integrals.
First, the resonance absorption density \cite{Pietralla1993}
\begin{align}
    \label{resonance_absorption_density}
    \alpha_{0 \to x, t} \left( E, z_t, ^A_Z\mathrm{X}\right) = & \sigma_{0 \to x, t} \left( E, ^A_Z\mathrm{X} \right) \frac{\Phi \left( E, z_t \right)}{\Phi \left( E, 0 \right)} \\
    \approx & \sigma_{0 \to x, t} \left( E, ^A_Z\mathrm{X} \right) \frac{\Phi \left( E, z_t \right)}{\Phi \left( E_x, 0 \right)} , \nonumber
\end{align}
was introduced.
In the second approximate equality of Eq.~\eqref{resonance_absorption_density}, the energy independence of the initial photon flux was used [Eq.~\eqref{constant_photon_flux}].
Second, the integral over the product of the detection efficiency and the angular correlation is denoted as a product of the target cross-section area $A_t$ and an equivalent mean value:
\begin{widetext}
\begin{equation}
    \label{efficiency_angular_correlation}
    A_t \langle \epsilon W \rangle_{0 \to x \to y, t} \left( E_x, z_t \right) = \iint_{A_t} \iint_{S_d} \epsilon \left(E_x, \mathbf{x}_d - \mathbf{x}_t \right) W_{0 \to x \to y} \left( \mathbf{x}_d - \mathbf{x}_t \right) \mathrm{d} x_t \mathrm{d} y_t.
\end{equation}
\end{widetext}

\noindent The impact of the $z_t$ dependence of the factor $\langle \epsilon W \rangle$ in Eq.~\eqref{efficiency_angular_correlation} was assessed by assuming that the efficiency depends on the direct path length $\Delta d \left(\mathbf{x}_d - \mathbf{x}_t \right)$ inside the target volume from the reaction vertex to the detector center, and the solid angle $\Omega \left( \mathbf{x}_t \right)$ at which the detector face is seen from the reaction vertex:
\begin{equation}
    \label{efficiency_approximation}
    \epsilon \left( E_x, \mathbf{x}_d - \mathbf{x}_t \right) \propto \exp \left[ -\mu_t \left( E_x \right) \Delta d \left( \mathbf{x}_d - \mathbf{x}_t \right) \right] \Omega \left( \mathbf{x}_t \right).
\end{equation}
As for Eq.~\eqref{constant_mass_attenuation_boron}, the mass attenuation coefficients of Hubbell and Seltzer \cite{HubbellSeltzer2004} were used to estimate the attenuation of the emitted gamma ray, while the approximation of Ref.~\cite{GardnerCarnesale1969} was employed for the solid angle.
The angular correlation was neglected, since the relevant transitions are either isotropic ($^6$Li) or their variation with solid angle is negligible ($^{11}$B).
\begin{figure}
    \includegraphics[width=0.45\textwidth, trim=50 50 40 50, clip]{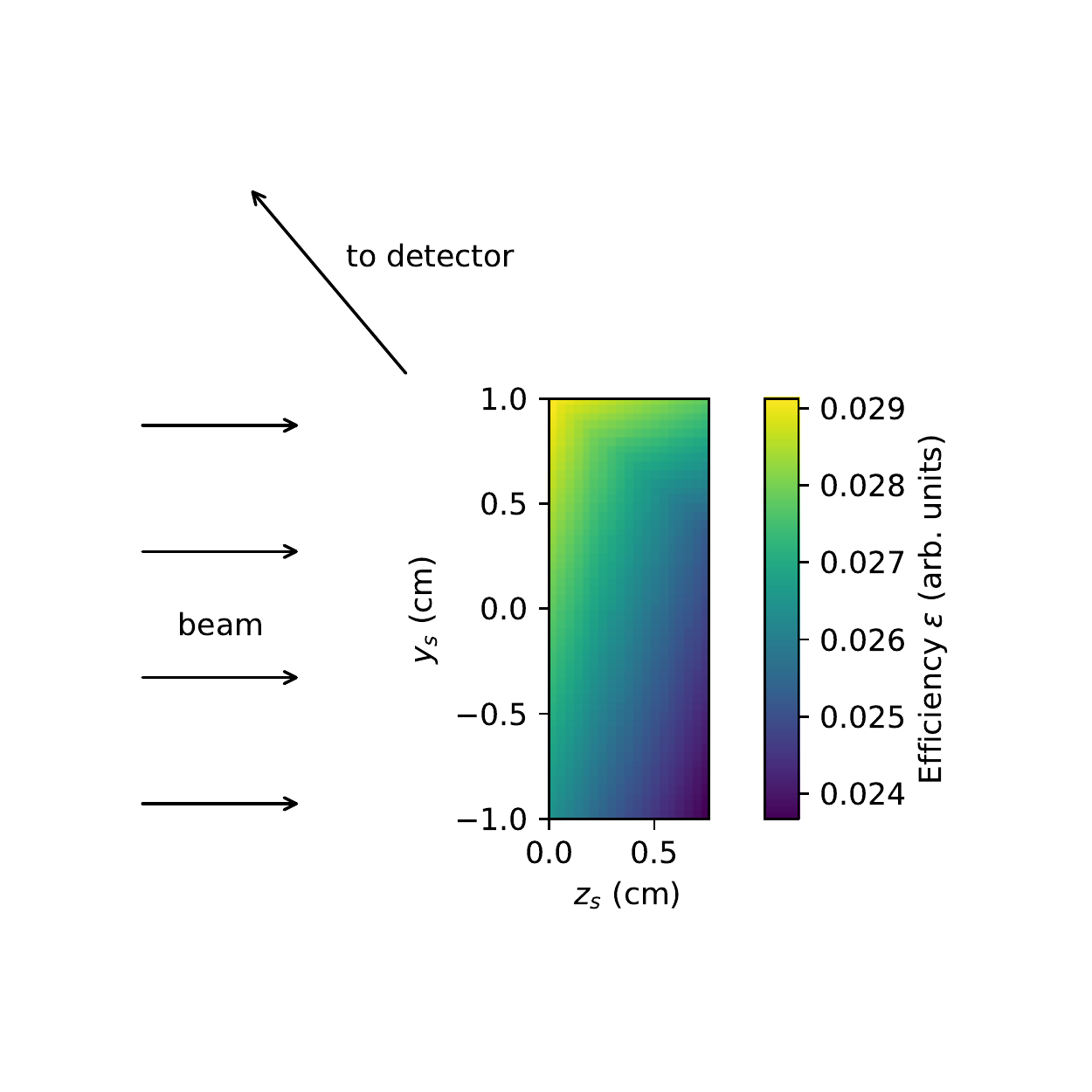}
    \caption{
        \label{efficiency_position_dependence} Cross section of the $^6$Li scatterer which shows the dependence of the detection efficiency of the $130^\circ$ detector on the reaction vertex position (central rectangle).
        The quantity $\epsilon \left( x_s = 0, y_s, z_s \right)$ was calculated using Eq.~\eqref{efficiency_approximation}.
        A legend to the right of the target indicates the color code.
        The direction of propagation of the incoming photon beam and the line of vision of the detector are indicated by arrows. 
    }
\end{figure}
As an example, Fig.~\ref{efficiency_position_dependence} shows $\epsilon \left( x_s = 0, y_s, z_s \right)$ for the scatterer and the detector at a polar angle of $130^\circ$.
For this detector, the efficiency changes by about 10\% between $z_{0,s}$ and $z_{1,s}$, which indicates a significant dependence of the experimental count rate on the dimensions and composition of the target.
However, the experimental self absorption $R_\mathrm{exp}$ is a ratio of expressions like Eq.~\eqref{nrf_reactions_energy_independence}.
Although the position dependence of $\langle \epsilon W \rangle$ cannot be factored out mathematically, it was found through numerical investigations that variations on the aforementioned order of magnitude have no impact on the result for $R_\mathrm{exp}$.
We anticipate this result here by approximating the product of the efficiency and the angular correlation by its value at the most upstream part of the target:
\begin{equation}
    \label{efficiency_angular_correlation_average}
    \langle \epsilon W \rangle_{0 \to x \to y, t} \left( E_x, z_t \right) \approx \langle \epsilon W \rangle_{0 \to x \to y, t} \left( E_x, z_{0,t} \right) 
\end{equation}
\begin{equation*}
    \mathrm{for}~z_t~\mathrm{in}~\left[ z_{0, t}, z_{1, t} \right],
\end{equation*}
so that Eq.~\eqref{nrf_reactions_energy_independence} becomes:
\begin{widetext}
\begin{align}
    \label{nrf_reactions_energy_and_position_independence}
    N_{0 \to x \to y, t} \left( ^A_Z \mathrm{X} \right) = \frac{\Gamma_{x \to y}}{\Gamma_x} n_t \left( ^A_Z \mathrm{X} \right) A_t \Phi \left( E_x, 0 \right) \langle \epsilon W \rangle_{0 \to x \to y, t} \left( E_x, z_{0,t} \right) \int_{E_x - \Delta E_x }^{E_x + \Delta E_x } \int_{z_{0, t}}^{z_{1, t}} \alpha_{0 \to x, t} \left( E, z_t, ^A_Z\mathrm{X} \right) \mathrm{d} z_t \mathrm{d} E.
\end{align}
\end{widetext}

\noindent At this point, it is instructive to consider the 'thin-target approximation'; i.e. the limiting case where the decay of the photon flux during the passage of the beam through the material is negligible:
\begin{equation}
    \label{thin_target_condition}
    \left[ \mu_t \left( E_x \right) + \sigma_{0 \to x, t} \left( E; ^A_Z\mathrm{X} \right) n_t \left( ^A_Z \mathrm{X} \right) \right] \left( z_{1, t} - z_{0, t} \right) \ll 1.
\end{equation}
The condition above concerns the argument of the exponential term in Eq.~\eqref{photon_flux}, which models the decay of the photon flux due to nonresonant scattering and the self-absorption effect as the beam passes through matter.
If the $z_t$ dependence of this term can be neglected, it is straightforward to show that Eq.~\eqref{nrf_reactions_energy_independence} reduces to
\begin{widetext}
\begin{equation}
    \label{nrf_reactions_thin_target}
    N_{0 \to x \to y} \left( ^A_Z\mathrm{X} \right) = \frac{\Gamma_{x \to y}}{\Gamma_x} \underbrace{n_t \left( ^A_Z \mathrm{X} \right) A_t \left( z_{1, t} - z_{0, t} \right)}_{N_t} \Phi \left( E_x, 0 \right) I_{0 \to x} \langle \epsilon W \rangle_{0 \to x \to y, t} \left( E_x \right),
\end{equation}
\end{widetext}
i.e. a product of the branching ratio, the number of target nuclei $N_t$, the photon flux at the beginning of the target, the integrated cross section, and the product of the efficiency and the angular correlation.

\noindent As described in the main text, the experiment consisted of a measurement with the scatterer and the two boron targets only~(`nrf'), and one with an additional absorber target~(`abs').
Using Eq.~\eqref{nrf_reactions_energy_and_position_independence} the unnormalized experimental self absorption $R_\mathrm{exp}^\prime$ is defined as:
\begin{widetext}
\begin{align}
    \label{unnormalized_self_absorption}
    R_\mathrm{exp}^\prime &= 1 - \frac{N_{1^+_1 \to 0^+_1 \to 1^+_1, s}^\mathrm{abs} \left( ^{6}\mathrm{Li} \right)}{N_{1^+_1 \to 0^+_1 \to 1^+_1, s}^\mathrm{nrf} \left( ^{6}\mathrm{Li} \right)} \nonumber \\
    &=1 - \frac{\Phi^\mathrm{abs} \left( E_{0^+_1}, 0 \right) \exp \left[ \mu_a \left( E_{0^+_1} \right) \left( z_{0, a} - z_{1, a} \right) \right]}{\Phi^\mathrm{nrf} \left( E_{0^+_1}, 0 \right) \phantom{\exp \left[ \mu_a \left( E_{0^+_1} \right) \left( z_{0, a} - z_{1, a} \right) \right]}} \\
    &\phantom{=1-} \times \frac{ \int_{E_{0^+_1} - \Delta E}^{E_{0^+_1} - \Delta E} \exp \left[ \sigma_{1^+_1 \to 0^+_1} \left( E; ^6\mathrm{Li} \right) \left( z_{0, a} - z_{1, a} \right) \right] \int_{z_{0, s}}^{z_{1, s}} \alpha_s \left( E, z_s; ^6\mathrm{Li} \right) \mathrm{d} z_s \mathrm{d} E}
    {\int_{E_{0^+_1} - \Delta E}^{E_{0^+_1} - \Delta E} \phantom{\exp \left[ \sigma_{1^+_1 \to 0^+_1} \left( E; ^6\mathrm{Li} \right) \left( z_{0, a} - z_{1, a} \right) \right]} \int_{z_{0, s}}^{z_{1, s}} \alpha_s \left( E, z_s; ^6\mathrm{Li} \right) \mathrm{d} z_s \mathrm{d} E}. \nonumber
\end{align}
\end{widetext}
Equation~\eqref{unnormalized_self_absorption} contains a ratio of double integrals that only depends on the absorption cross section, but not on the absolute scale of the scatterer dimensions and its nonresonant scattering cross section.
However, the ratio of the photon flux in the two measurements and the nonresonant attenuation in the absorber remain and prevent a model-independent extraction of the cross section.
They were eliminated here by considering the ratio of count rates for the resonances of the normalization targets:
\begin{widetext}
\begin{align}
    \label{normalization_target_ratio}
    f\left( E_x \right) = \frac{N_{3/2^- \to x \to y, u}^\mathrm{abs} \left( ^{11}\mathrm{B} \right) + N_{3/2^- \to x \to y, d}^\mathrm{abs} \left( ^{11}\mathrm{B} \right)}{N_{3/2^- \to x \to y, u}^\mathrm{nrf} \left( ^{11}\mathrm{B} \right) + N_{3/2^- \to x \to y, d}^\mathrm{nrf} \left( ^{11}\mathrm{B} \right)} = 
    \frac{\Phi^\mathrm{abs} \left( E_x, 0 \right) \exp \left[ \mu_a \left( E_x \right) \left( z_{0, a} - z_{1, a} \right) \right]}{\Phi^\mathrm{nrf} \left( E_x, 0 \right) \phantom{\exp \left[ \mu_a \left( E_x \right) \left( z_{0, a} - z_{1, a} \right) \right]}}.
\end{align}
\end{widetext}
The factor $f$ has a well-known smooth energy dependence \cite{Schiff1951, HubbellSeltzer2004}, which can be calculated with high accuracy.
This was validated in the present experiment by an additional off-beam measurement of the attenuation coefficient of the absorber with a $^{56}$Co source  \cite{Romig15b}.
Consequently, it can be interpolated to $E_{0^+_1}$ to obtain the relative self absorption
\begin{equation}
    \label{relative_self_absorption}
    R_\mathrm{exp} = 1 - \frac{1}{f\left( E_{0^+_1} \right)} \frac{N_{1^+_1 \to 0^+_1 \to 1^+_1, s}^\mathrm{abs} \left( ^{6}\mathrm{Li} \right)}{N_{1^+_1 \to 0^+_1 \to 1^+_1, s}^\mathrm{nrf} \left( ^{6}\mathrm{Li} \right)}.
\end{equation}
Equation~(1) in the main text is a symbolic abbreviation of Eq.~\eqref{relative_self_absorption}.
\subsection{Note on energy-dependent quantities}

\noindent In the previous section, it was argued that the detection efficiency $\epsilon$ [Eq.~\eqref{constant_efficiency}], the mass attenuation coefficient $\mu$ [Eq.~\eqref{constant_mass_attenuation}], and the photon flux $\Phi$ [Eq.~\eqref{constant_photon_flux}] can be considered to be energy-independent within the narrow range determined by the width of the resonance, at the precision level of the present experiment.
The approximations were made without taking into account the energy dependence of the resonance cross section.
However, if the resonance is symmetric, i.e.,
\begin{equation}
    \label{symmetric_cross_section}
    \sigma \left( E_x - \Delta E \right) = \sigma \left( E_x + \Delta E \right)
\end{equation}
\begin{equation*}
    \mathrm{for}~\Delta E~\mathrm{in}~[ 0, \infty ),
\end{equation*}
the results from above are still valid if the quantity $X$, which may be any of $\left\{ \epsilon, \mu, \Phi \right\}$, varies linearly with the energy.
In the following, the Taylor expansion around the resonance energy $E_x$ is used:
\begin{equation}
    \label{taylor_expansion}
    X \left( E \right) = X \left( E_x \right) + \frac{\mathrm{d} X}{\mathrm{d} E}\Big|_{E = E_x} \left( E - E_x \right) + \mathcal{O} \left( E^2 \right).
\end{equation}
The expression for the number of resonantly scattered photons $N_{0 \to x \to y}$ [see the general expression Eq.~\eqref{nrf_reactions_general} or Eqs.~\eqref{nrf_reactions_z_axis}, \eqref{nrf_reactions_energy_independence}, and \eqref{nrf_reactions_energy_and_position_independence}] contains an energy integral over a product of $X$ and $\sigma$:
\begin{align}
    &\phantom{=} \int_{E_x - \Delta E}^{E_x + \Delta E} X \left( E \right) \sigma \left( E \right) \mathrm{d} E \\
    &\approx \int_{E_x - \Delta E}^{E_x + \Delta E} \left[ X \left( E_x \right) + \frac{\mathrm{d} X}{\mathrm{d} E}\Big|_{E = E_x} \left( E - E_x \right) \right] \sigma \left( E \right) \mathrm{d} E \nonumber \\
    &=X \left( E_x \right) I. \nonumber
\end{align}
In the last equality, the energy-integrated cross section [Eq.~\eqref{integrated_cross_section}] has been inserted.
The term proportional to the first derivative of $X$ vanishes due to the symmetry properties of $\sigma$ [Eq.~\eqref{symmetric_cross_section}] and the expression $E - E_x$.

\noindent A Doppler-broadened Breit-Wigner cross section with the parameters of the present analysis fulfils the requirements of this section to a good approximation, i.e., even a linear variation of $\epsilon$, $\mu$, or $\Phi$ with the energy can be ignored.
Therefore, the order-of magnitude estimates of the systematic uncertainty given above can be considered to be conservative.
\section{Data processing}
\noindent Details about the processing of the raw spectra can be found in Ref.~\cite{Romig15b}.
Here, it will be assumed that the number of resonantly scattered photons $N_{0 \to x \to y}$, which follows a Poisson distribution (see, e.g., Ref.~\cite{Zyla2020}), has already been determined for all relevant transitions in both measurements.
These include the resonance of interest in $^6$Li, and the ground-state transitions of the excited states of $^{11}$B at \unit[2125]{keV}, \unit[4445]{keV}, and \unit[5020]{keV}.

\begin{figure}
    \includegraphics[width=0.5\textwidth, trim=0 0 20 0, clip]{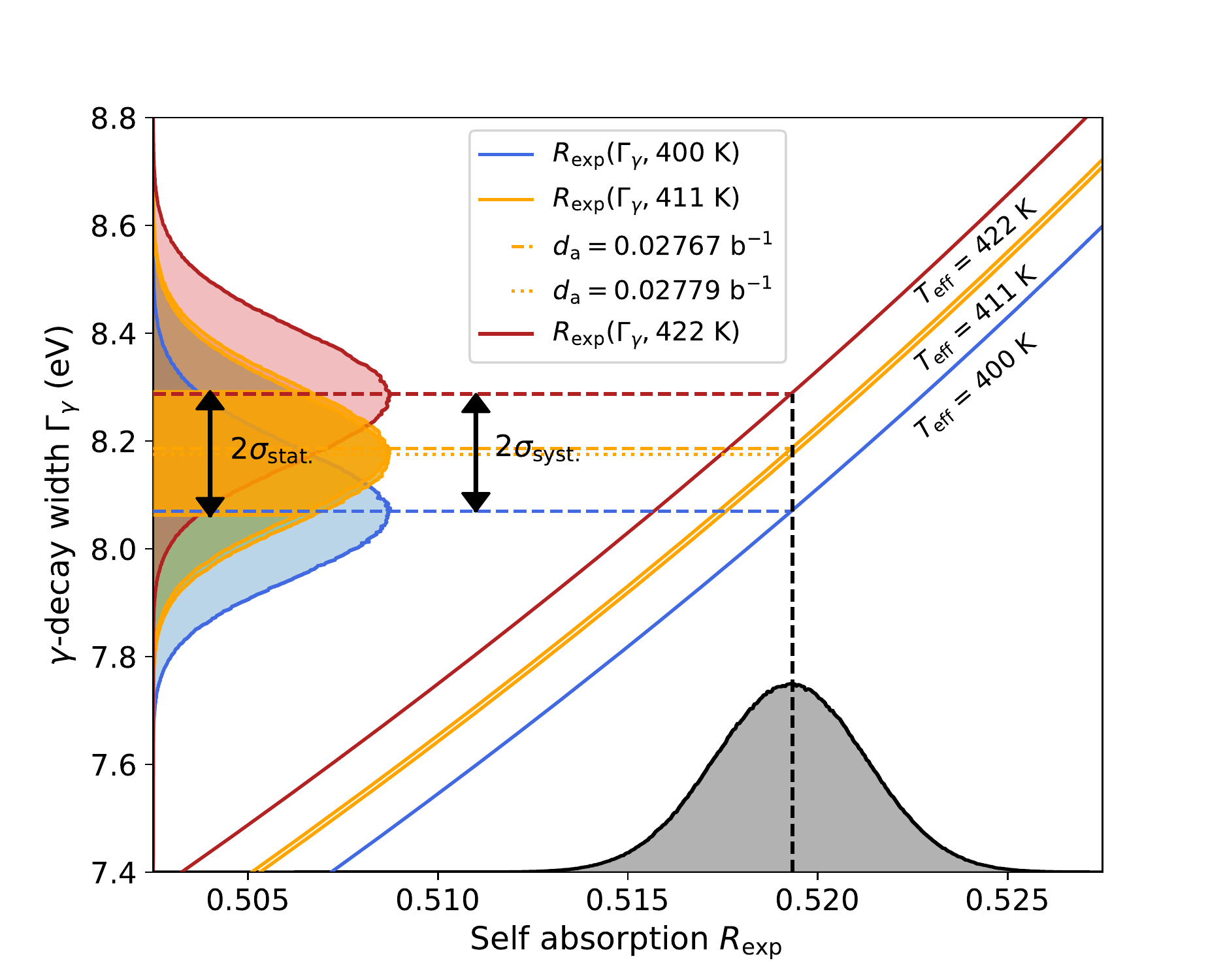}
    \caption{
        \label{R_to_Gamma} Visualization of the different contributions to the total uncertainty of $\Gamma_\gamma$.
        A probability distribution for $\Gamma_\gamma$ is obtained from a projection of the distribution of $R$ (abscissa) on $\Gamma_\gamma$ (ordinate) via the unique functional relation $\Gamma_\gamma \left( R \right)$.
        Four different dependencies of $\Gamma_\gamma$ on $R$ using different combinations of $T_\mathrm{eff}$ and $d_\mathrm{abs}$ are presented.
        The color code indicates whether the lower limit (blue), the upper limit (red), or the mean value (orange) of $T_\mathrm{eff}$ was used.
        For the mean value of the temperature, two possible distributions are shown.
        They correspond to the extrema of the \unit[68]{\%} CI of $d_\mathrm{abs}$, while the other two temperatures used the mode of $d_\mathrm{abs}$.
        The resulting distributions for $\Gamma_\gamma$ are shown in the same color as the respective curve.
        Dashed and dotted horizontal and vertical lines indicate the projection of the mode of $R$ on the mode of $\Gamma_\gamma$.
        A filled area in the distributions for $T_\mathrm{eff} = \unit[411]{K}$ indicates the \unit[68]{\%} CI of the statistical uncertainty (including the uncertainty of $d_\mathrm{abs}$), while the difference between the modes of the distributions for $T_\mathrm{eff} = \unit[400]{K}$ and $T_\mathrm{eff} = \unit[422]{K}$ represents the total systematic uncertainty.
    }
\end{figure}

\noindent The correction factor $f \left( E_{0^+_1} \right)$, according to Eq.~\eqref{relative_self_absorption}, was obtained by scaling a simulated \cite{Agostinelli03, Allison06, Allison16} curve $f \left( E \right)$ to the experimental count-rate ratios of the $^{11}$B transitions.
Due to the small relative uncertainties of the count rates, on the order of less than one percent, a $\chi^2$ minimization procedure was used, which resulted in a normally-distributed value of $f \left( E_{0^+_1} \right) = 1.631(6)$.
Since the aforementioned notch-refilling effect due to small-angle scattering inside the collimator artificially increases the count rate in the measurement with the absorber, the total correction factor in the present analysis amounted to \cite{Romig15b}:
\begin{equation}
    \label{total_correction_factor}
    f^\prime \left( E_{0^+_1} \right) = \frac{1}{1.0033} f \left( E_{0^+_1} \right).
\end{equation}
With the correction factor included, and all approximations of the previous section applied, the self-absorption coefficient $R_\mathrm{exp}$ was treated as a function of the $\gamma$-decay width $\Gamma_\gamma$, the effective temperature $T_\mathrm{eff}$, and the thickness of the absorber target $d_\mathrm{a} = z_{a, 1} - z_{a, 0}$:
\begin{equation}
    R_\mathrm{exp} = R_\mathrm{exp} \left( \Gamma_\gamma, T_\mathrm{eff}, d_\mathrm{a} \right).
\end{equation}
As described in the main text, two different calculations of $T_\mathrm{eff}$ were interpreted as a lower and upper limit for the effective temperature, and they are denoted as a systematic uncertainty $\sigma_\mathrm{syst}$ here.
The absorber thickness contributes to the systematic uncertainty as well, and it was assumed to follow a normal distribution.
Using a Monte-Carlo method described in the `Guide to the expression of uncertainty in measurement' \cite{GUM2008}, the uncertainty given in the main text was obtained.
The strictly monotone relation between $R_\mathrm{exp}$ and $\Gamma_\gamma$ allows for a straightforward visualization of all contributions to the total uncertainty (Fig.~\ref{R_to_Gamma}).
\providecommand{\noopsort}[1]{}\providecommand{\singleletter}[1]{#1}%